\documentclass[reprint,amsmath,aps,pra,footinbib,notitlepage,longbibliography,superscriptaddress]{revtex4-1}
\usepackage{graphicx,epstopdf,dcolumn,bm,mathtools,siunitx,multirow,hyperref,xcolor,amsfonts,physics}

\begin{document}
\title{Coherent charge carrier dynamics in the presence of thermal lattice vibrations}

\author{Donghwan Kim}
\affiliation{Department of Chemistry and Chemical Biology, Harvard University, Cambridge,
Massachusetts 02138, USA}

\author{Alhun Aydin}
\affiliation{Department of Chemistry and Chemical Biology, Harvard University, Cambridge,
Massachusetts 02138, USA}
\affiliation{Department of Physics, Harvard University, Harvard University, Cambridge, Massachusetts 02138, USA}

\author{Alvar Daza}
\affiliation{Department of Physics, Harvard University, Harvard University, Cambridge, Massachusetts 02138, USA}
\affiliation{Nonlinear Dynamics, Chaos and Complex Systems Group, Departamento de  F\'isica, Universidad Rey Juan Carlos,
Tulip\'an s/n, 28933 M\'ostoles, Madrid, Spain}

\author{Kobra N.~Avanaki}
\affiliation{Department of Chemistry and Chemical Biology, Harvard University, Cambridge,
Massachusetts 02138, USA}
\affiliation{Department of Physics, Harvard University, Harvard University, Cambridge, Massachusetts 02138, USA}

\author{Joonas Keski-Rahkonen}
\affiliation{Department of Chemistry and Chemical Biology, Harvard University, Cambridge,
Massachusetts 02138, USA}
\affiliation{Department of Physics, Harvard University, Harvard University, Cambridge, Massachusetts 02138, USA}

\author{Eric J. Heller}
\email{eheller@fas.harvard.edu}
\affiliation{Department of Chemistry and Chemical Biology, Harvard University, Cambridge,
Massachusetts 02138, USA}
\affiliation{Department of Physics, Harvard University, Harvard University, Cambridge, Massachusetts 02138, USA}

\date{\today}
\begin{abstract}
We develop the coherent state representation of lattice vibrations to describe their interactions with charge carriers. In direct analogy to quantum optics, the coherent state representation leads from quantized lattice vibrations (phonons)  naturally to a quasiclassical field limit,  i.e., the deformation potential.
To an electron, the deformation field is a sea of  hills and valleys, as ``real'' as any external field, morphing and propagating  at the sound speed, and growing in magnitude with temperature.
In this disordered potential landscape, the charge carrier dynamics is treated nonperturbatively, preserving their coherence beyond single collision events. We show the coherent state picture agrees exactly with the conventional Fock state picture in perturbation theory. Furthermore, it goes beyond by revealing aspects that the conventional theory could not explain: transient localization even at high-temperatures by charge carrier coherence effects, and band tails in the density of states due to the self-generated disorder (deformation) potential in a pure crystal. The coherent state paradigm of lattice vibrations supplies tools for probing important questions in condensed matter physics as in quantum optics. 
\end{abstract}
\maketitle
\section{Introduction}\label{s:intro}

Crystal lattice vibrations had initially been treated essentially as a classical field, but in the early papers one phonon perturbation theory was adopted~\cite{ShockleyBardeen,BS2}.
Since the introduction of second quantization,  lattice vibrations have been treated as a quantum field in the Fock state picture. In this description,
the particle (phonon) aspect of lattice vibrations gets the most attention by design, and one rarely if ever thinks of the lattice vibrations as a classical field.

There is a viable alternative to the Fock state description, just as in quantum optics: coherent states are equivalent yet permit the construction of the quasiclassical field from the quantum field of lattice vibrations.
The wave aspect of  lattice vibrations is thus emphasized, providing a different perspective from the usual Fock state picture.
Although it is common in quantum optics to use coherent states to describe quantized electromagnetic wave~\cite{Sudarshan1963,glauber1963}, coherent states have rarely been used in condensed matter physics to describe quantized lattice vibrations.

In the conventional Fock state description, the interaction of an electron with lattice vibrations requires a phonon creation or annihilation within  first-order perturbation theory. Higher-order interactions are approximated as an incoherent and uncorrelated chain of the first-order events through Boltzmann transport theory~\cite{ashcroft1976solid,abrik,ziman2001electrons,Many-Particle}. In this way, any electron coherence lasting from one collision to the next has been  neglected. Given that a phonon bath   has changed in the host lattice, it might seem reasonable to neglect coherence, by the usual bath-induced decoherence arguments. However, in the quantum optical analog, this argument would be equivalent to saying that electrons cannot behave coherently in a strong electromagnetic field, which is of course not true.

The conventional Bloch-Gr{\"u}neisen theory~\cite{Bloch1930,Grueneisen1933} employing the Fock states picture describes the temperature dependence of  electrical resistivity of most metals by taking the scattering of electrons from acoustic phonons into account.
However, despite the success of the theory, there are major phenomena where the conventional methods seem to fall short, such as  the mystifying linear temperature dependence of resistivity and universal scattering rate in strange metals~\cite{strangemetal1,Legros2018,Zaanen2019,Varma2020,Greene2020,planckianmetals, Tlinearresistivity}.
The perspective in this work, using a coherent state description of lattice vibrations, may shed light on these unexplained phenomena.


In the coherent state representation of the lattice vibrations, we lose both the ability and the need to count  phonons. This is replaced by information about the phase and amplitude of each vibrational normal mode, leaving the occupation numbers  uncertain. Here, we explore the overlooked part of the wave-particle duality for lattices, considering lattice vibrations to be \emph{waves}, instead of particles (phonons).   We will refrain from using the word \textit{phonon} and instead use \textit{lattice vibrations} if possible to emphasize the wave nature of quantum lattice vibration. A phonon, after all, is a single and countable quantum, a particle like a photon. 

There are two different applications of the term \textit{coherence} used here. First, we preserve \textit{electron coherence}  over multiple scatterings, and second, there is the \textit{coherence of lattice vibrations} described by coherent states. 

\begin{figure*}[t]
\centering
\includegraphics[width=0.95\textwidth]{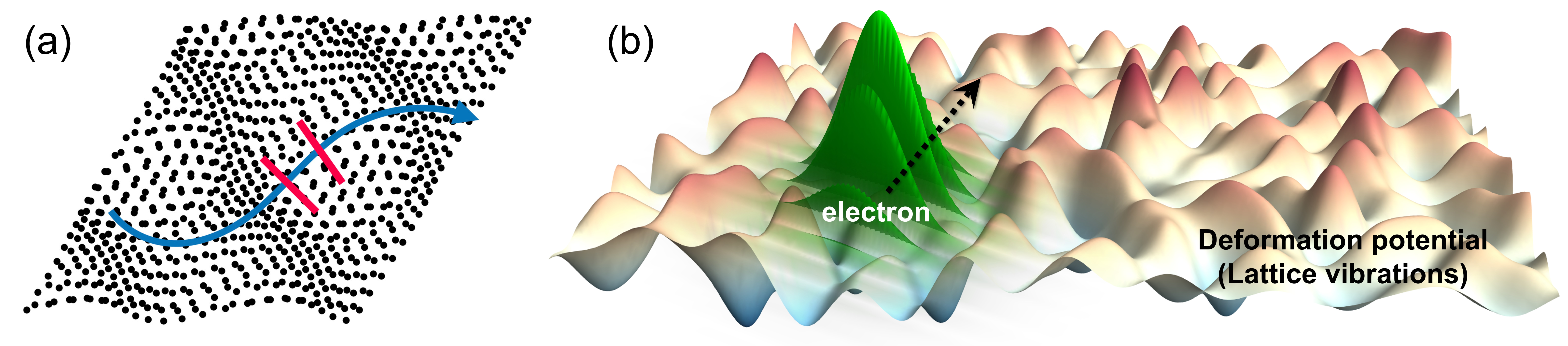}
\caption{(a) Schematic of an atomic lattice subject to acoustic deformations. Classical path of an electron subject to the resulting deformation potential is shown by blue arrow. (b) A particular realization of the coherent state lattice vibrations at a certain temperature. Electron wave packet (real part shown in green and its direction denoted by dotted black arrow) coherently propagates in a spatially continuous internal field formed by the acoustic deformations. Electrons quasielastically scatter (quite similar to impurity scattering) from the disordered landscape formed by the lattice vibrations.}
\label{Deformation_potential_example}
\end{figure*}

We derive a quasiclassical field, called the deformation potential, from the quantum field for the interaction of an electron with lattice vibrations within the coherent state representation.
The deformation potential is schematically shown in Fig. \ref{Deformation_potential_example}.
The interactions of an electron with this field are quasielastic and mostly remain coherent.
Such electron coherence is  absent in the conventional theories of electron-phonon interaction such as the Bloch--Gr\"{u}neisen theory.
Below, we do find  agreement with the conventional theory whenever electron coherence is not important.  This work is therefore viewed as an extension of the current theory into the coherent regime, agreeing with the conventional approaches in the normal regimes. The electron coherence shows interesting physics such as temporary (or transient) localization, even at high-temperatures, and existence of band tails in density of states.

The paper is organized as follows. In Sec.~\ref{s:historical}, we provide a historical background indicating the explicit and implicit use of the coherent state description of the lattice vibrations and electromagnetic waves in literature.
In Sec.~\ref{s:CoherentState}, we introduce the definition and properties of a coherent state, and then consider its application to the lattice vibrations by discussing its advantages over a Fock state.
In Sec.~\ref{s:Deformation}, we derive the explicit form of the quasiclassical field of the deformation potential and discuss its properties.
In Sec.~\ref{s:comparison}, we calculate transport scattering rate in the perturbation theory in coherent state picture and show its equivalence to the Fock state picture.
In addition, the differences between quantum and classical fields are discussed.
In Secs. \ref{s:ElecDynamics} and \ref{s:coherence}, we perform full quantum calculations using the split operator method.
We also implemented semiclassical ray trajectory calculations in Appendix \ref{app:SemiclassicalRay}. These are well justified in some but certainly not all regimes of temperature and doping. In Sec.~\ref{s:ElecDynamics}, we construct a  temperature--Fermi momentum phase diagram from quantum dynamics simulations of an electron in the deformation potential.
We go beyond recovering the conventional theory in Sec.~\ref{s:coherence}, demonstrating the existence and consequence of electron coherence and multiple scattering effects which were neglected before. In particular, we show that electrons ``attempt'' to localize  in a short time, and band tails in density of states are caused by the disorder (deformation) potential.
We discuss the possible implications of our findings and conclude in Sec.~\ref{s:conclusion}.

\section{Historical background}\label{s:historical}


The concept of what we now refer to as the coherent state of a harmonic oscillator was  introduced by Schr{\"o}dinger in 1926. Employing his time-dependent equation, he showed that a displaced ground state oscillates without changing shape, with the mean position and mean momentum  obeying  classical equations of motion~\cite{schr}. The extension of this result to many oscillators, including   harmonic solids, is direct and straightforward~\cite{hellerkim}. This shows that any  classical behavior like sound propagating through a harmonic lattice
has an exact quantum analog within a coherent state representation.
  
Coherent states appeared  in a different context in 1954, beginning with Hanbury Brown and Twiss's~\cite{HBT1,HBT2}     interferometric measurement of   apparent stellar diameters using two telescope mirrors spaced a variable distance apart. The field arriving   from a distant   star is extremely weak and presumably incoherent, so that arriving quanta (particle picture)  at   distant detectors were expected to be uncorrelated. Instead, Hanbury Brown and Twiss found the signal to be correlated, like advancing waves would be at nearby points on a beach. The correlation degrades as the telescopes are moved farther apart, and the decorrelation distance reveals the apparent diameter of the distant star.  The implication that such weak light from an incoherent source arrived as waves did not receive a warm welcome, well after the discovery of the photoelectric effect.  Eventually, however, the battle of wave {\it vs.} particle paradigms regarding the Hanbury Brown and Twiss controversy initiated the unification of the two paradigms in 1963, sparking the field of  quantum optics, with coherent states playing a central role~\cite{Sudarshan1963,glauber1963,scully1997quantum}.

Nowadays, in addition to quantum optics~\cite{walls2007quantum, scully1997quantum, grynberg2010introduction, gerry2005introductory}, coherent states play an important role, e.g., in studies of light-matter interaction in cavity quantum electrodynamics~\cite{Walther2006}, quantum chaos such as scarring~\cite{PhysRevB.96.094204.2017, JPhysCondensMatter.31.105301.2019, PhysRevLett.123.214101.2019}, novel states of quantum matter such as superconductivity or superfluidity~\cite{tinkham2004introduction}, and quantum fields in general~\cite{lancaster2014quantum}. Despite the triumph of the coherent state picture in physics, its advantage in describing  the dynamics of a lattice has remained elusive, with a few rare exceptions indicating the possibility of an unused asset, such as Refs.~\cite{noolandi_use_1972,van_kranendonk,hellerkim,Heller2021}.  
In this paper, we want to amend this conceptual shortcoming. 

This work may be viewed as a recapitulation of the Hanbury Brown--Twiss story: what had always been treated as particles (phonons in Fock states picture) is sometimes better viewed as waves (coherent states picture), within the context of a unified wave-particle theory. A terahertz (THz) lattice mode has about 13 quanta at 100~K, and a gigahertz (GHz) mode has 13 000.  However, no matter how few quanta occupy the modes in a Fock space, the coherent state picture is valid, although it may not necessarily be in the classical limit. In any case, weakly occupied modes play a minor role in electrical resistivity.
We also remark that even a field equivalent to one photon arriving per second would show the wave-like Hanbury Brown--Twiss interference effect.

The traditional approach to the interaction of an electron with lattice vibrations traces back to the 1950 paper of Bardeen and Shockley~\cite{ShockleyBardeen,BS2} who introduced the notion of a  deformation potential experienced by electrons, resulting from the acoustic wave compression and dilation of the lattice. There was a moment when a classical field picture could have been adopted,  but the deformation potential has instead ever since been employed exclusively in Fock states description with the first-order perturbation theory~\cite{Khan1984,Pipa2001,Zgaren2014,SantiagoPerez2015,SantiagoPerez2017,Liu2017,RevModPhys2017}. 

The traces of the idea of employing deformation potential as a real, nonperturbative  field can be found only on  dusty shelves of the literature. A prescient suggestion in this direction was made as early as in 1959 by Holstein in a footnote~\cite{holstein_theory_1959}, where he suggested a treatment of phonons based on the classical lattice-vibration wave packets, implying that he suspected the advantages of treating electrons as evolving nonperturbatively in a classical (not quantized) lattice field. However, the promised work never materialized~\cite{privcom}.

Whenever lattice vibrations (such as sound waves) are treated classically, the coherent state representation is implied: the amplitude and phase of an oscillation give the coherent state, specifying the positions and momenta of all the atoms in the lattice. For example, Pippard~\cite{pippard_theory_1960} treated the ultrasound field produced by a transducer as a classical wave in order to interpret the observed rapid attenuation of ultrasound in metals.  It would indeed seem strange to utilize a second quantized occupation-number formalism for a classically occupied mode with billions of quanta.  

In the classic solid state textbook by Ashcroft and Mermin~\cite{ashcroft1976solid}, two chapters are devoted to ``semiclassical" methods, by which is meant treating external fields acting on electrons in metals as classical fields, where the kinetic part of the Hamiltonian is governed by the band structure. Lifschitz and Kosevich~\cite{lifkos} took this methodology further by developing a coherent, semiclassical analysis, revealing the contribution of coherent semiclassical orbits on the Fermi surface to magnetic field oscillations, including the Shubnikov--de Haas effect~\cite{LifshitsEM1958TotS}. It is evident that a coherent quantum treatment of the conduction electron in external (or internal as we do here) fields is necessary for interference effects like Shubnikov--de Haas. Non-equilibrium Green's functions (NEGF) formalism also provides a solid framework for the study of quantum dynamics and coherence effects in general~\cite{dattabook1995,dattabook2005}. While NEGF is used for coherence-preserving elastic scattering due to impurities, the scattering of electrons due to lattice vibrations have almost exclusively been employed as an incoherent process, unlike what we do in this paper.

Over the years, there have been other developments and suggestions related to what we propose here. For example, in his text \textit{Solid State Theory}~\cite{harrison1980solid}, Harrison wrote the following:
\begin{quote}
``Because of the low frequencies of the acoustical modes, it is possible to correctly compute their contributions to the electron scattering by conceptually freezing the atoms at their positions in the deformed crystal and computing the electron scattering associated with the corresponding distortions. {\it Just as we calculated the scattering by defects in crystals}'' (italics ours).
\end{quote}
 
``Freezing" the atoms in position is very far from a Fock state.  It is  closer to a coherent state representation. Thus, it has been suspected for a long time that the deformation field can be taken at face value as a potential which would scatter electrons at the correct rate,  acting like a sea of defects in the process. This is also implied by the formal equivalence of the occupation number and coherent state representations~\cite{Loudon:105699}, which is discussed further in Sec. \ref{s:comparison}.  
 
This ``elastic" program is already highly developed in the field of thermal diffuse scattering (TDS) from crystals. In one version of TDS, collimated electron pulses are sent through crystals; both Bragg and diffuse scattering result. The diffuse scattering increases with temperature, and changes with time if vibrational population evolution is occurring. The frozen lattice (adiabatic) approximation, i.e., supposing the lattice to be fixed at typical configurations as the electrons pass through, works extremely well at explaining the diffuse scattering and better, quantitatively exploiting it for the inverse scattering problem, yielding the geometry of lattice vibrational modes and vibrational energy evolution~\cite{bosakbetween, siwick}.

Nevertheless, the explicit utilization of the coherent state representation for lattice motion is uncommon in the literature. A notable exception is the 1972 paper by Noolandi and Kranendonk (see Refs.~\cite{noolandi_use_1972,van_kranendonk}). In their work, the aim was mainly to understand solid hydrogen, but  the paper has been largely ignored. It is difficult to find other examples taking a conceptual step towards this direction, except some of our own related works~\cite{hellerkim,Heller2021,mohanty_lazy_2019}.  
\section{Coherent state description of lattice vibrations}\label{s:CoherentState}
In this section, we introduce the definition and basic properties of a coherent state and its utility for describing the lattice vibrations. The coherent state picture is an  appropriate way to treat the lattice vibrations fully quantum mechanically, of equal, unassailable stature to the conventional Fock (number) state approach relying on the concept of individual, particlelike phonons.

A classical field consists of waves with well-defined amplitudes and phases. However, this is not the case when the field is  quantum mechanical.   Fluctuations are associated with the amplitudes and phases. A field in a number state $\vert n \rangle$ has a well-defined amplitude, but  lacks knowledge of   phase. A field defined by a coherent state has an equal amount of uncertainty in both amplitude and phase. The field phase $\varphi$ and particle number $n$ satisfy the   uncertainty principle~\cite{PhysRevA.48.3159.1993, PhysRevA.39.1665.1989, JModOpt.36.7.1989}:
\begin{equation}\label{uncertainty_principle}
     \Delta n  \Delta \varphi \gtrsim 1.
\end{equation}
The coherent state satisfies $\Delta n =  \Delta \varphi$, and it is a special case of a more general class of states which may have reduced uncertainty in one parameter at the expense of increased uncertainty in the other. Such states are  known as squeezed states~\cite{walls_squeezed_1983}, e.g., an amplitude squeezed state ($\Delta n < 1$) or a phase squeezed state ($\Delta \varphi < 1$). These states can be created from coherent states by employing a unitary squeeze operator.

\begin{figure}[t]
\centering
\includegraphics[width=0.40\textwidth]{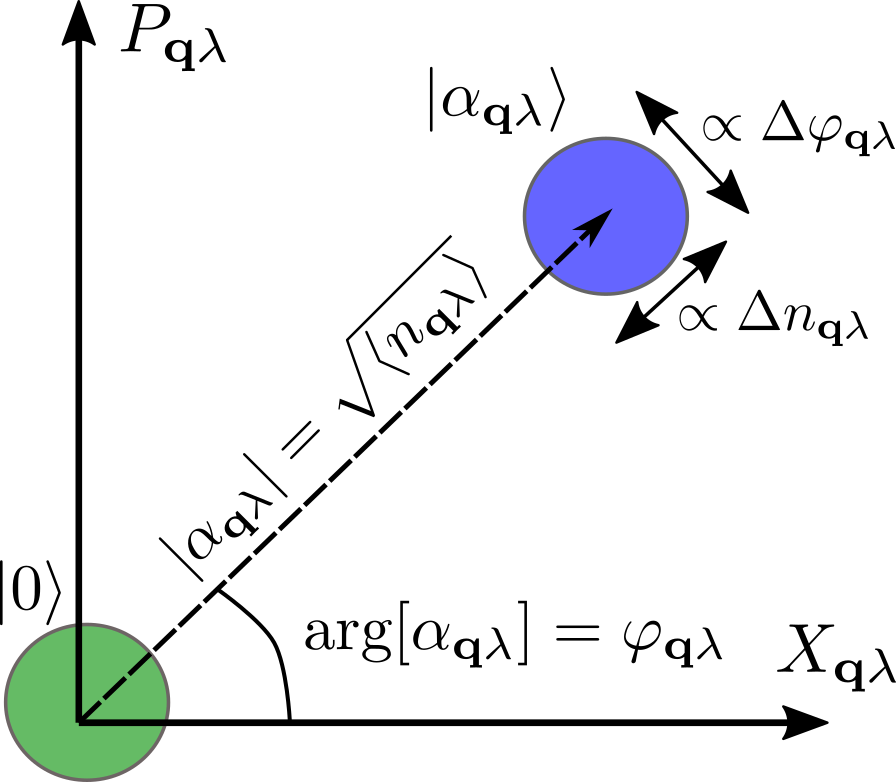}
\caption{Representation of a coherent state as a displaced form (blue circle) of vacuum state (green circle) in the phase space of a (dimensionless) normal coordinate $X_{\mathbf{q}\lambda}=(\hat{a}_{\mathbf{q}\lambda}^\dag+\hat{a}_{\mathbf{q}\lambda})/2$ and its conjugate momentum $P_{\mathbf{q}\lambda}=(\hat{a}_{\mathbf{q}\lambda}^\dag-\hat{a}_{\mathbf{q}\lambda})i/2$. The coherent state $\vert \alpha_{\mathbf{q}\lambda} \rangle$ is identical to the vacuum state $\vert 0 \rangle$ except the coordinate shift with an amplitude $\sqrt{\langle n_{\mathbf{q}\lambda}\rangle}$ and the rotation by a phase $\varphi_{\mathbf{q}\lambda}$. Thus, the coherent state is identified with the complex parameter $\alpha_{\mathbf{q}\lambda} = \sqrt{\langle n_{\mathbf{q}\lambda}\rangle}\exp(i\varphi_{\mathbf{q}\lambda})$ with quantum fluctuations in the amplitude $\Delta n_{\mathbf{q} \lambda}$ and in the phase $\Delta \varphi_{\mathbf{q} \lambda}$, which are restricted by the uncertainty principle. In the case of a coherent state, the amplitude and the phase have equal dispersion $\Delta n_{\mathbf{q} \lambda}=\Delta \varphi_{\mathbf{q} \lambda}$, in contrast to a more generic squeezed state.}
\label{fig:coherent_state}
\end{figure}

In this paper, we associate each normal mode with a wave vector $\mathbf{q}$ and a branch index $\lambda$ of the lattice vibration to a coherent state $\vert \alpha_{\mathbf{q}\lambda} \rangle$ where $\alpha_{\mathbf{q}\lambda}$ is a complex parameter. These states are most easily defined via the unitary displacement operator composed of the annihilation $\hat{a}_{\mathbf{q}\lambda}$ and creation operators $\hat{a}_{\mathbf{q}\lambda}^{\dagger}$ of the field as 
\begin{equation}
    \mathcal{D}(\alpha_{\mathbf{q}\lambda}) = e^{ \alpha_{\mathbf{q}\lambda}\hat{a}_{\mathbf{q}\lambda}^\dag-\alpha_{\mathbf{q}\lambda}^*\hat{a}_{\mathbf{q}\lambda} }.
    \label{eq:displOp}
\end{equation}
With the displacement operator, the coherent state is generated as 
\begin{equation}\label{coherent_state}
    \vert \alpha_{\mathbf{q}\lambda} \rangle = \mathcal{D}(\alpha_{\mathbf{q}\lambda}) \vert 0 \rangle,
\end{equation}
where $\vert 0 \rangle$ is the vacuum (ground) state. This construction coincides with the original definition of being an eigenstate of the annihilation operator, i.e.,
\begin{equation}
    \hat{a}_{\mathbf{q}\lambda} \vert \alpha_{\mathbf{q}\lambda} \rangle = \alpha_{\mathbf{q}\lambda} \vert \alpha_{\mathbf{q}\lambda} \rangle.
\end{equation}
By definition, a coherent state $\vert \alpha_{\mathbf{q}\lambda} \rangle$ contains an indefinite number of quanta, but it has an average number of quanta of
\begin{equation}
    \langle n_{\mathbf{q}\lambda}\rangle = \langle \alpha_{\mathbf{q}\lambda} \vert \hat{a}_{\mathbf{q}\lambda}^{\dagger} \hat{a}_{\mathbf{q}\lambda} \vert \alpha_{\mathbf{q}\lambda} \rangle =  \vert \alpha_{\mathbf{q}\lambda} \vert^2.
    \label{eq:avenummode}
\end{equation}
As illustrated in Fig.~\ref{fig:coherent_state}, the coherent state $\vert \alpha_{\mathbf{q}\lambda} \rangle$ can be interpreted as the displaced form of the vacuum state $\vert 0 \rangle$, and the state is characterized by its amplitude $\sim\abs{\alpha_{\mathbf{q}\lambda}}=\sqrt{\langle n_{\mathbf{q}\lambda}\rangle}$ and phase $\arg(\alpha_{\mathbf{q}\lambda})=\varphi_{\mathbf{q} \lambda}$, whose fluctuations are bounded by the uncertainty principle in Eq.~\eqref{uncertainty_principle}.
Even though these coherent states are not orthogonal, they form an overcomplete basis for the corresponding Hilbert space. A more detailed discussion on the coherent states can be found, e.g., in Refs.~\cite{walls2007quantum, scully1997quantum, grynberg2010introduction, gerry2005introductory}.

The coherent state picture offers an alternative description to conventionally used number states: The latter emphasizes the particle nature of the lattice vibrations, whereas the former accentuates the wave nature.
However, this wave-particle duality is normally hidden by the approximations the two limits encourage. Nonetheless, although at the most fundamental level  the two pictures are equivalent, there are two clear benefits of employing the coherent state over the number-state representation.

The first virtue of coherent states is that they are the closest quantum mechanical states to a classical description allowed by the uncertainty principle. This quantum-classical correspondence enables us to construct \emph{quasiclassical} fields from quantum fields, and to study the boundary between the classical and quantum realms in general, reflecting Schr{\"o}dinger's original idea of coherent states~\cite{schr}. In the limit of macroscopic occupation, the coherent state picture blends into the concept of a classical field with a fixed amplitude and phase. It is also as close as possible to a specification of the instantaneous positions and momenta of the atoms in the lattice. This explains why coherent states $\vert \alpha_{\mathbf{q} \lambda} \rangle$ are referred to as quasiclassical. On the downside, because they are intrinsically in motion, like the lattice itself, the lattice coherent states ultimately demand a time dependent description of the field they generate on the electron.  Today, this is not much of a barrier to implementation. 

Second and more importantly, the coherent states $\vert \alpha_{\mathbf{q}\lambda} \rangle$ are \emph{robust} against the effect of the environment. In fact, these states are pointer states, i.e., they correspond to some value of a pointer in a classical measuring apparatus (see, e.g., Refs.~\cite{wheeler2014quantum, RevModPhys.75.715.2003, PhysRevD.24.1516.1981}). In other words, the pointer in the measuring device can only have classical probability. Any coherent superposition of states $\vert \alpha_{\mathbf{q}\lambda} \rangle$ is fragile and it will rapidly decay to a classical probability distribution of different coherent states~\cite{wheeler2014quantum, RevModPhys.75.715.2003, PhysRevD.24.1516.1981, joos2003decoherence,schlosshauer2007decoherence}. For example, a cat state constructed of coherent states will break down into a classical probability distribution since the external interactions easily destroy the quantum coherence of the initial state~\cite{haroche2006exploring, PhysRevA.49.2785.1994, PhysRevA.50.2548.1994, PhysRevA.49.1266.1994}. In a similar manner, squeezed states are also fragile against an influence of environment (see, e.g., Refs.~\cite{PhysRevD.47.488.1993, PhysRevA.67.022107.2003}). In particular, in the presence of a perturbation such as dissipation, a high-number Fock state $\vert n_{\mathbf{q}\lambda} \rangle$ promptly decomposes into an incoherent linear combination of coherent states $\vert \alpha_{\mathbf{q}\lambda} \rangle$ with different phases $\varphi_{\mathbf{q}\lambda}$ and amplitude $\sim \sqrt{n_{\mathbf{q}\lambda}}$ (see, e.g., Refs.~\cite{grynberg2010introduction, JOptB.4.5418.2002, OptExpress.2.131.2001,PhysRevLett.101.240401.2008}). In this light, it is natural to describe the lattice vibrations in terms of the coherent states.

Since the vibrational normal modes are independent of one another, the entire lattice vibrations as a whole can be described with the product state of the coherent states $\ket{\alpha_{\mathbf{q}\lambda}}$ of the normal modes $\mathbf{q}\lambda$'s, a multimode coherent state
\begin{equation}
    \ket{\bm{\alpha}}=\prod_{\mathbf{q}\lambda}\ket{\alpha_{\mathbf{q}\lambda}},
    \label{eq:Vecalpha}
\end{equation}
as considered in Ref.~\cite{hellerkim}, where $\bm{\alpha}=(\dots,\alpha_{\mathbf{q}\lambda},\dots)$ is the collection of the complex numbers $\alpha_{\mathbf{q}\lambda}$ of the normal modes. A realistic lattice is coupled to an environment which is expressible employing thermal ensembles of coherent states~\cite{thermalcoherent, scully}. Therefore, we consider that each mode is in thermal equilibrium with a heat bath at temperature $T$, and the average number of quanta of corresponding coherent states follows the Bose--Einstein statistics, i.e.,
\begin{equation}
    N_{\mathbf{q}\lambda} =\langle n_{\mathbf{q}\lambda}\rangle_{\textrm{th}} = \frac{1}{\exp\left(\hbar\omega_{\mathbf{q}\lambda}/k_{\textrm{B}} T\right) - 1} .
    \label{eq:BEoccup}
\end{equation}
The amplitudes of the modes  ${\sim}\abs{\alpha_{\mathbf{q}\lambda}}$ are determined by taking the thermal average value in Eq. \eqref{eq:BEoccup} for the average occupation in Eq. \eqref{eq:avenummode}:
\begin{align}
    \abs{\alpha_{\mathbf{q}\lambda}}^2=N_{\mathbf{q}\lambda}.
    \label{eq:ThermalAvg}
\end{align}
This approach of associating the thermal average value with the average occupation has been widely utilized as a natural pathway to thermalize a field, e.g., by Bardeen and Shockley for a classical field of lattice vibration~\cite{BS2}, and by Hanbury Brown and Twiss for a classical field of electromagnetic wave~\cite{HBT1}. 

In general, the coherent state $\vert \bm{\alpha} \rangle$ gives the exact quantum dynamics of the entire lattice. However, we can take a next step by following the similar road as in the theory of quantized electromagnetic fields: by merging the concepts of the quantum lattice vibration field and the (thermal) coherent states, we construct a quasiclassical lattice vibration field that yields the \emph{deformation potential}. 

\section{Deformation Potential}\label{s:Deformation}

Inspired by the coherent state picture of lattice vibrations described in the previous section, we introduce a framework for the interaction of an electron with lattice vibrations. We begin with second-quantized form of  lattice vibrations, then derive the deformation potential employing the thermal ensemble of coherent states.

The concept of the deformation potential was first introduced by Bardeen and Shockley for nonpolar semiconductors~\cite{ShockleyBardeen,BS2}. The main idea is that local electronic band energy can be used as an effective potential when the variation of the lattice distortions is sufficiently gradual.
This is the case for the long-wavelength longitudinal acoustic lattice deformation which interacts with electrons of thermal velocity.

In this work, we focus on the acoustic lattice deformation since its thermal population is much larger than the optical lattice deformation. Nonetheless, if necessary, optical deformation potential can also be considered and employed in a similar manner to the acoustic deformation potential~\cite{Many-Particle, PhysRev.104.331.1956, PhysRev.104.649.1956, PhysRev.104.1281.1956, PhysRevB.24.2025.1981, PhysRevB.104.195201.2021}.


\subsection{Derivation}

In a distorted lattice, lattice deformation is characterized by a displacement field $\mathbf{u}(\mathbf{x})$ that  is a displacement of an atom at a position $\mathbf{x}$ from its equilibrium position.
The displacement field gives strain fields $\epsilon_{ij}(\mathbf{x})=\pdv{u_i(\mathbf{x})}{x_j}$ ($u_i$ and $x_j$ are $i$th and $j$th components of $\mathbf{u}$ and $\mathbf{x}$, respectively). The local electronic band energy $E\left(\mathbf{k};\epsilon_{ij}(\mathbf{r})\right)$ is a function of strain fields $\epsilon_{ij}(\mathbf{r})$ evaluated at the electron position $\mathbf{r}$. For simplicity, we assume the lattice has a cubic (or square in two dimensions) symmetry; we consider a more generic case in Appendix \ref{app:GeneralDeriv}.

Assuming the amplitudes of the strain fields are small, we can expand the conduction (valence) band energy in the strain fields about the equilibrium lattice configuration
\begin{align}
    E\left(\mathbf{k};\epsilon_{ij}(\mathbf{r})\right)
    &=
    E_0(\mathbf{k})
    +
    E_d\nabla\cdot\mathbf{u}(\mathbf{r})
    +
    \cdots,\label{eq:DP2}
\end{align}
where $\mathbf{k}$ is an electron wave vector, $E_0(\mathbf{k})$ is band energy of undistorted (periodic) lattice, $E_d$ is the deformation potential constant, and $\nabla\cdot\mathbf{u}(\mathbf{r})$ is dilation. The first-order correction term in the expansion defines a \textit{deformation potential}:
\begin{align}
    V_D(\mathbf{r})=E_d\nabla\cdot\mathbf{u}(\mathbf{r}).
    \label{eq:DPdef}
\end{align}
In principle, the deformation potential constant $E_d$ has $\mathbf{k}$-dependence, but it is usually taken to be a constant~\cite{BS2}. The material-dependent $E_d$ value is determined either experimentally~\cite{DPconstCu,DPconstCu2,DPconstSi,DPconstSi2} or computationally, e.g., employing the density functional theory~\cite{PhysRevB.104.195201.2021}.

Next we quantize the displacement field $\mathbf{u}(\mathbf{x})$ in a similar fashion as in quantum field theory.
The displacement field operator $\hat{\mathbf{u}}(\mathbf{x},t)$ is expressed through the creation $a_{\mathbf{q}\lambda}^{\dagger}$ and annihilation $a_{\mathbf{q}\lambda}$ operators of modes identified with wave vector $\mathbf{q}$, branch index $\lambda$, and angular frequency $\omega_{\mathbf{q}\lambda}$. Here $\lambda$ actually refers to the polarization index of acoustic modes since we only focus on acoustic lattice deformations. In this mode expansion, the quantum displacement field is given as
\begin{eqnarray}\label{displament_field}
\begin{aligned}
    \hat{\mathbf{u}}(\mathbf{x},t)
    &=
    i\sum_{\mathbf{q},\lambda}\sqrt{\frac{\hbar}{2\rho_m \mathcal{V}\omega_{\mathbf{q}\lambda}}}
    {\bm \varepsilon}_{\mathbf{q}\lambda}
    \\
    &\qquad\qquad\quad\times
    \left( a_{\mathbf{q}\lambda}e^{-i\omega_{\mathbf{q}\lambda}t}
    +
    a_{-\mathbf{q}\lambda}^\dag e^{i\omega_{\mathbf{q}\lambda}t} \right)
    e^{i\mathbf{q}\cdot\mathbf{x}},
    \label{eq:uquantum}
\end{aligned}
\end{eqnarray}
where ${\bm \varepsilon}_{\mathbf{q}\lambda}$ is the polarization unit vector, $\rho_m$ and $\mathcal{V}$ are the mass density and the volume (or the area in two dimensions) of the solid, respectively. The field operator above can be understood as a canonically quantized version of the classical lattice displacement field considered, e.g., in Refs.~\cite{landau2013classical, Hamaguchi2017}. It also agrees with the quantum field presented in the textbook by Mahan [see Eq. (1.85) of Ref. \cite{Many-Particle}].

Analogous to Eq. \eqref{eq:DPdef}, the quantum field of the deformation potential is determined as
\begin{eqnarray}
\begin{aligned}
    \hat{V}_D(\mathbf{r},t)
    &=
    E_d\nabla\cdot\hat{\mathbf{u}}(\mathbf{r},t)
    \\
    &=
    -\sum_{\mathbf{q}}
    g_{\mathbf{q}l}
    (a_{\mathbf{q}l}e^{-i\omega_{\mathbf{q}l}t}
    +
    a_{-\mathbf{q}l}^\dag e^{i\omega_{\mathbf{q}l}t})
    e^{i\mathbf{q}\cdot\mathbf{r}},
    \label{VDquantumfield}
\end{aligned}
\end{eqnarray}
where the parameter
\begin{equation}
    g_{\mathbf{q}l}=E_d
\sqrt{\frac{\hbar}{2\rho_m \mathcal{V}\omega_{\mathbf{q}l}}}
\abs{\mathbf{q}}
\label{ephcoupling}
\end{equation}
represents electron-phonon coupling strength. It should be noted that only the longitudinal ($\lambda=l$) acoustic modes contribute to the deformation potential $(\mathbf{q}\cdot\bm{\varepsilon}_{\mathbf{q}l}=\abs{\mathbf{q}})$, whereas the transverse ($\lambda=t$) acoustic modes do not $(\mathbf{q}\cdot\bm{\varepsilon}_{\mathbf{q}t}=0)$.

We next construct a corresponding \emph{quasiclassical} field of the deformation potential $V_D(\mathbf{r},t)$ by taking the expectation value of the quantum field of the deformation potential $\hat{V}_D(\mathbf{r},t)$ with respect to the multimode coherent state $\ket{\bm{\alpha}}$ [see Eq.~\ref{eq:Vecalpha}] describing the lattice:
\begin{eqnarray}\label{quasiclassical_DP}
\begin{aligned}
    V_D(\mathbf{r},t)
    &=\bra{\bm{\alpha}}\hat{V}_D(\mathbf{r},t)\ket{\bm{\alpha}}
    \\
    &=
    -\sum_{\mathbf{q}}
    g_{\mathbf{q}l}
    (\alpha_{\mathbf{q}l}e^{-i\omega_{\mathbf{q}l}t}
    +
    \alpha_{-\mathbf{q}l}^* e^{i\omega_{\mathbf{q}l}t})
    e^{i\mathbf{q}\cdot\mathbf{r}}.
\end{aligned}
\end{eqnarray}
The amplitudes of the modes  ${\sim}\abs{\alpha_{\mathbf{q}\lambda}}$ are determined by the thermal occupations [see Eq. \eqref{eq:ThermalAvg}].

Now we use the Debye model that introduces linear dispersion $\omega_{\mathbf{q}l}=v_s|\mathbf{q}|$ ($v_s$ is sound speed) and Debye wave number (isotropic cutoff) $q_D$.
Then, the quasiclassical field of the deformation potential is written as
\begin{align}
    V_D(\mathbf{r},t)
    &=
    -\sum_{\substack{\mathbf{q}\\q<q_D}}
    2g_{\mathbf{q}l}
    \sqrt{N_{\mathbf{q}l}}
    \cos(\mathbf{q}\cdot\mathbf{r}-\omega_{\mathbf{q}l}t+\varphi_{\mathbf{q}l}),
    \label{eq:VDcl}
\end{align}
where $\varphi_{\mathbf{q}\lambda}=\arg(\alpha_{\mathbf{q}\lambda})$ is the phase of a coherent state $\ket{\alpha_{\mathbf{q}\lambda}}$.
Examples of this potential are given in Figs. \ref{Deformation_potential_example}, \ref{Deformation_potential_temperature_dependence}, and \ref{ACDP}.

We want to emphasize two aspects regarding the derivation.
First, although our deformation potential was derived from the quantized lattice vibrations, it can also be deduced from the classical lattice vibrations as shown in Appendix~\ref{app:ClassicalDeriv}. Second, the lattice model is not necessarily
 the Debye model and any appropriate dispersion relation $\omega_{\mathbf{q}l}$ can be used. Nevertheless, we will employ the Debye model, assuming the linear dispersion $\omega_{\mathbf{q}l} = v_s \vert \mathbf{q} \vert$ where $v_s$ is sound velocity.


We should also note that a very similar expression  was derived in 1957 by Hanbury Brown and Twiss for the vector potential of a blackbody field~\cite{HBT1}. The forces acting on electrons in the blackbody field are very similar to the forces acting on electrons in the deformation potential field. However, the essential difference between these two is in the existence of the ultraviolet cutoff $q_D$ in the deformation potential originating from the minimal lattice spacing.

\subsection{Properties}\label{s:DPproperties}
The derived quasiclassical deformation potential field has important statistical properties. We treat phases $\{\varphi_{\mathbf{q}l}\}_{\mathbf{q}}$ appearing in Eq. \eqref{eq:VDcl} as uniformly distributed random variables.
Then, the deformation potential value $V_D$
is a random variable normally distributed with mean $\mu_{V_D}$ and standard deviation $\sigma_{V_D}$ by the central limit theorem, i.e.,
\begin{align*}
    V_D\sim\mathcal{N}(\mu_{V_D},\sigma_{V_D}^2).
\end{align*}
The mean is zero $\mu_{V_D}=\ev{V_D}=0$ and the standard deviation is root-mean-square of the potential values $\sigma_{V_D}=\sqrt{\ev{V_D^2}}=V_{\mathrm{rms}}$.
Note the average $\ev{\cdot}$ can be taken over either position $\mathbf{r}$, time $t$, or phase $\varphi_{\mathbf{q}l}$; they all give the identical results as they appear in the argument of the same cosine in Eq. \eqref{eq:VDcl} \cite{randvars}.

The potential is homogeneously random~\cite{zimandisorder} in space and in time, meaning the probability distribution of $V_D$ does not depend on a position $\mathbf{r}$ or time $t$ given that the phases $\varphi_{\mathbf{q}l}$ are random variables.
Thus, each spatiotemporal section of the potential is statistically indistinguishable from another.

The spatiotemporal autocorrelation function $C(\bm{\delta}\mathbf{r},\delta t)=\ev{V_D(\mathbf{r},t)V_D(\mathbf{r}+\bm{\delta}\mathbf{r},t+\delta t)}$ gives the strength of the potential fluctuation and the decay of its spatiotemporal correlation (see more details in Appendix \ref{app:autocorr}). In two dimensions (2D), the autocorrelation is
\begin{align}
    C^{\mathrm{(2D)}}(\delta r,\delta t)
    &=
    \frac{\mathcal{V}}{(2\pi)^2}
    \int_0^{q_D}
    g_{\mathbf{q}l}^2
    N_{\mathbf{q}l}
    \pi J_0(q\delta r)
    \cos(v_sq\delta t)q\dd q.
    \label{eq:AC2D}
\end{align}
The autocorrelation is significant for a spatiotemporal relation $\delta r=v_s\delta t$ corresponding to the sound-wave propagation.

From the autocorrelation, we can obtain the typical energy scale of the potential fluctuation, i.e., the root mean square of the potential values $V_{\textrm{rms}}=\sqrt{C(0,0)}$.
Note although the electron-phonon coupling strength of each mode $g_{\mathbf{q}l}\sim1/\sqrt{\mathcal{V}}$ has a volume dependence, the potential fluctuation $V_{\textrm{rms}}$ does not. This is because the number of modes ${\sim}\mathcal{V}$ cancel the volume dependence out as shown in $\mathcal{V}g_{\mathbf{q}l}^2$ factor in Eq. \eqref{eq:AC2D}.

The typical length scale of the potential is determined by its largest wave-number components. 
At temperature $T$, a Bose wave number $q_B(T)=k_BT/\hbar v_s$ determines the effective largest wave number from the thermal occupation. The modes below the Bose wave number [$q\lesssim q_B(T)$] are thermally active while the modes above the Bose wave number [$q\gtrsim q_B(T)$] are effectively frozen out; the Bose wave number $q_B(T)$ acts as a soft thermal cutoff. The actual effective thermal cutoff is given roughly by $5q_B(T)$ as the factor $g_{\mathbf{q}l}\sqrt{N_{\mathbf{q}l}}$ in Eq. \eqref{eq:VDcl} becomes negligible for $q\gtrsim 5q_B(T)$.
Thus, the effective wave number cutoff is the minimum of $5q_B(T)$ and $q_D$, i.e., $q_{\mathrm{eff}}(T)=\min\{5q_B(T),q_D\}$, which determines the length scale of the potential.

\begin{figure}[t]
\centering
\includegraphics[width=0.49\textwidth]{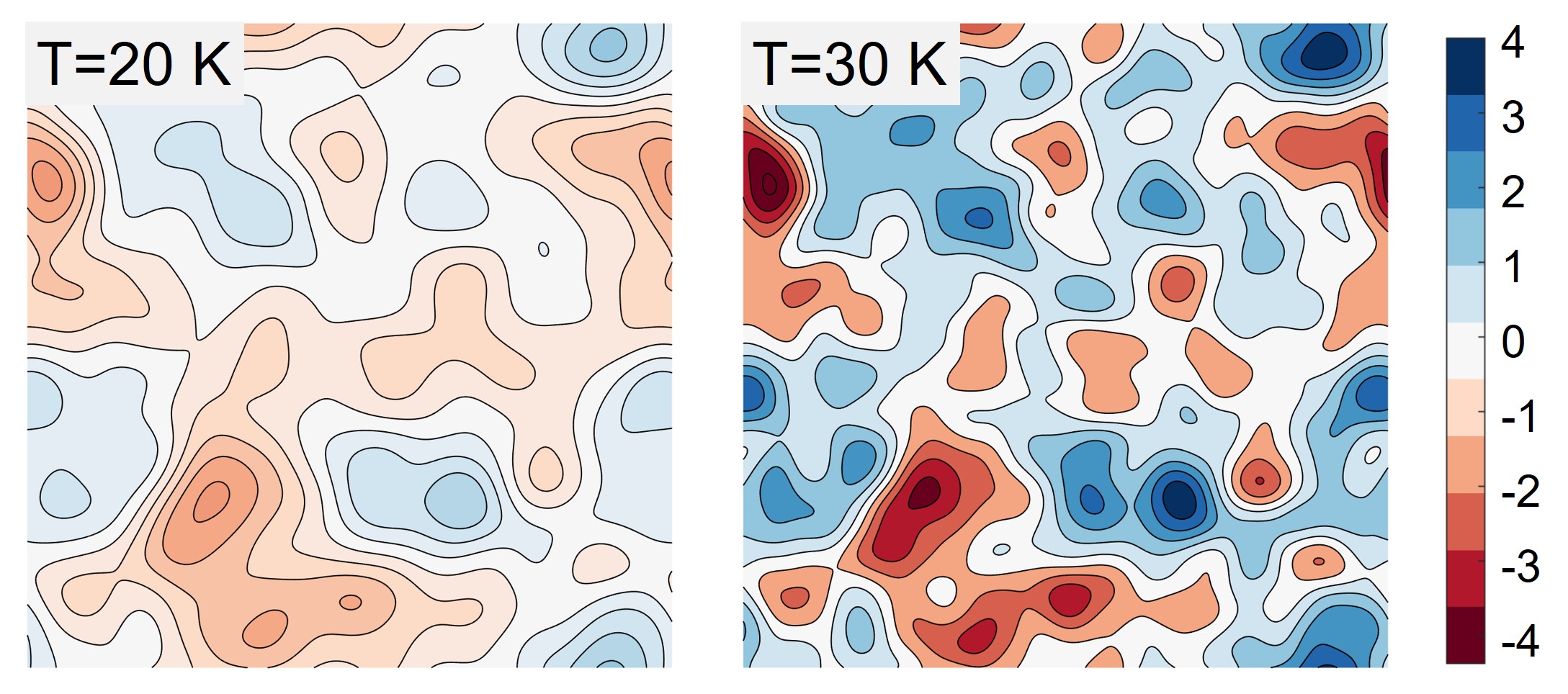}
\caption{Contour plots of the deformation potential in real space at $20$ and $\SI{30}{K}$ demonstrating the opening of vibrational modes with increasing temperature. The bumps and dips also get higher and deeper with increasing temperature. The legend and axes are given in arbitrary scale for illustration purposes. Identical random phases are used to generate the deformation potential for both temperatures. }
\label{Deformation_potential_temperature_dependence}
\end{figure}

For $T<0.2T_D$, the potential effectively does not ``notice'' the existence of the Debye cutoff $q_D$ since the thermal cutoff $5q_B(T)$ comes first, i.e., $q_{\mathrm{eff}}(T)=5q_B(T)$.
Figure \ref{Deformation_potential_temperature_dependence} shows the deformation potential at the two different temperatures both below $0.2T_D$. In this temperature range, as the temperature $T$ increases, the effective largest wave number $q_{\mathrm{eff}}(T)=5q_B(T)$ increases, thereby shorter length scale emerges.
In addition, as the amplitude of each mode increases as the temperature increases, the bumps and dips get higher and deeper.

\begin{figure}
    \centering
    \includegraphics[width=.40\textwidth]{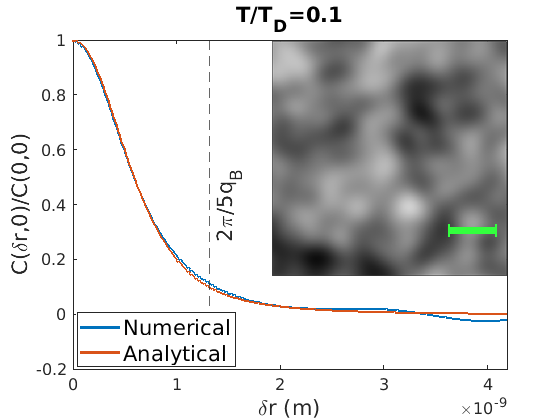}
    \includegraphics[width=.40\textwidth]{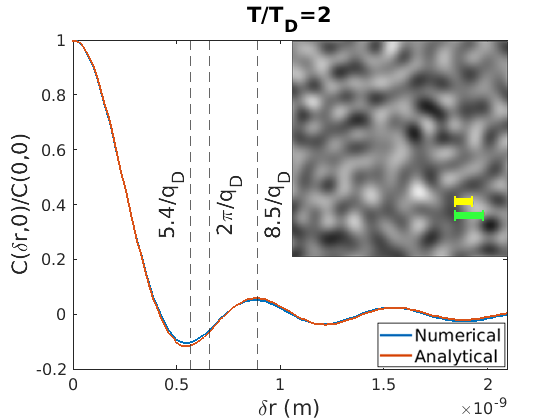}
    \caption{Potentials (insets) and their spatial autocorrelation functions at the two different temperatures $T=0.1T_D$ (top panel) and $T=2T_D$ (bottom panel).
    The blue curve (``Numerical'') is from the numerical evaluation of the autocorrelation $C(\bm{\delta}\mathbf{r},0)=\ev{V_D(\mathbf{r},t)V_D(\mathbf{r}+\bm{\delta}\mathbf{r},t)}$ for one realization of the deformation potential, and the red curve (``Analytical'') is from the evaluation of the analytical expression \eqref{eq:AC2D}.
    At $T=0.1T_D$, $2\pi/5q_B$ is the length scale of the autocorrelation decay.
    The sizes of the black dips and white bumps are similar to the length $2\pi/5q_B$ of the green stick.
    At $T=2T_D$, the autocorrelation has first negative and positive peaks at $5.4/q_D$ and $8.5/q_D$, respectively; both are in a similar order as the characteristic length scale of the cutoff $2\pi/q_D$.
    As a result, the potential shows granular structure which was not seen in the temperature $T=0.1T_D$.
    In the potential, the distance between a white bump and an adjacent black dip indeed matches with the length $5.4/q_D$ of the yellow stick. Also, the distance between adjacent white bumps (or black dips) agrees with the length $8.5/q_D$ of the green stick.}
    \label{ACDP}
\end{figure}

Furthermore, as shown in  the top panel of Fig. \ref{ACDP}, the sizes of bumps and dips are characterized by the length scale $2\pi/5q_B$ of the spatial autocorrelation decay.
For $T> 0.2T_D$, the potential notices the Debye cutoff $q_D$, i.e., $q_{\mathrm{eff}}(T)=q_D$ and the relevant length scale $\sim2\pi/q_D$ is seen in both the potential and the autocorrelation function as shown in the bottom panel of Fig. \ref{ACDP}.

Similarly, the typical timescale of the potential change is determined by its largest frequency components.
Temporal autocorrelation function $C(0,\delta t)$ gives the characteristic timescale of the change of the potential, which is given by $2\pi/5\omega_B(T)$ for $T<0.2T_D$, and $2\pi/\omega_D$ for $T>0.2T_D$  where $\omega_B(T)=v_sq_B(T)$ and $\omega_D=v_sq_D$ (see more details in Appendix \ref{app:autocorr}).
\subsection{Usage}
The electronic band energy in Eq. \eqref{eq:DP2} can be treated as a Hamiltonian for an electron:
\begin{align}
    H(\hbar\mathbf{k},\mathbf{r},t)
    &=
    E_0(\mathbf{k})
    +
    V_D(\mathbf{r},t)
    \label{DPHam}
\end{align}
where $\mathbf{k}$ is an electron wave vector, $E_0(\mathbf{k})$ is     band energy of undistorted (hence periodic) lattice, and $V_D(\mathbf{r},t)$ is the quasiclassical field of the deformation potential in Eq. \eqref{eq:VDcl}.
The dynamics of the electron under the Hamiltonian
can be studied by solving the time-dependent Schr\"{o}dinger equation.

We use the effective mass model  $E_0(\mathbf{k})=\hbar^2k^2/2m^*$ where $m^*$ is an effective mass of an electron.
Any initial wave function can be used, but the Gaussian wave packet with an initial momentum is a reasonable choice for studying electron transport.
Although any energy of conduction (valence) electron can be considered, we mostly examine electrons with Fermi energy $E_F=\hbar^2k_F^2/2m^*$ as the important carriers.
 Electron wave packets with an initial momentum $\hbar k_F$ are thus investigated.

When the Fermi velocity is far faster than the sound speed,  $v_F\gg v_s$, the electron quickly enters into a new region that is uncorrelated to its original region.
Then, as the potential is homogeneously random, the uncorrelated new region is statistically indistinguishable from another region of the potential at another time.
Thus, the electron dynamics in the deformation potential $V_D(\mathbf{r},t)$ can be approximated as the dynamics in a frozen deformation potential $V_D(\mathbf{r},t=0)$ when $v_F\gg v_s$.
In this paper, we mainly use the frozen deformation potential, and the time dependence of the potential is discussed when it is necessary.

The Drude theory of metals states that, for a metal with a carrier of an effective mass $m^*$, an absolute value of charge $e$, and carrier density $n$, electrical resistivity $\rho$ is determined by the momentum relaxation time $\tau$ through the relation $\rho=m^*/ne^2\tau$.
 We focus on the inverse momentum relaxation time $1/\tau$ in place of resistivity $\rho$.

\section{Perturbation Theory in Coherent state picture}\label{s:comparison}

We calculate transport scattering rate in the perturbation theory in coherent state picture and show its equivalence to the Fock state picture.
In addition, the differences between quantum and classical fields are discussed.
We encourage readers to look at Appendix \ref{app:PertCohFock} for details of the calculations.

\subsection{Equivalence of coherent and Fock state descriptions in perturbation theory}\label{s:EquivCohFock}
Consider a system of a harmonic solid with electrons and quantized lattice vibrations. For simplicity, we think of a single electron Hamiltonian and contemplate Fermi statistics afterwards. Then, the system Hamiltonian is written as
\begin{align*}
    \hat{H}=\hat{H}_0+\hat{V}
\end{align*}
where
\begin{align*}
    \hat{H}_0
    =
    \frac{\hat{\mathbf{p}}^2}{2m^*}
    +
    \sum_{\mathbf{q}}\hbar\omega_{\mathbf{q}l}(a_{\mathbf{q}l}^\dag a_{\mathbf{q}l}+1/2)
\end{align*}
is the sum of the electronic kinetic energy and the elastic energy of lattice vibrations and 
\begin{align}
    \hat{V}
    =
    \sum_{\mathbf{q}}
    g_{\mathbf{q}l}
    (a_{\mathbf{q}l}
    +
    a_{-\mathbf{q}l}^\dag)
    e^{i\mathbf{q}\cdot\mathbf{r}},
    \label{eq:Vhat:main}
\end{align}
is their interaction energy in the Schr\"{o}dinger picture [cf. Eq. \eqref{VDquantumfield}].

Let us describe the lattice with a multimode coherent state $\ket{\bm\alpha}$ defined in Eq. \eqref{eq:Vecalpha}.
Then, consider a scattering of an initial many-body state $\ket{\mathbf{k},\bm\alpha}$, where $\mathbf{k}$ is an electron wave vector, by the quantum deformation field $\hat{V}$.
Time-dependent perturbation theory treats $\hat{H}_0$ as the unperturbed Hamiltonian and $\hat{V}$ as a perturbation \cite{sakurai2011modern}.
Then, the transport scattering rate (or the inverse of the momentum relaxation time) can be calculated from the momentum autocorrelation
\begin{align}
    C(t)
    &=
    \bra{\mathbf{k},\bm\alpha}\frac{\hat{\mathbf{p}}\cdot\hat{\mathbf{p}}(t)+\hat{\mathbf{p}}(t)\cdot\hat{\mathbf{p}}}{2}\ket{\mathbf{k},\bm\alpha}\notag
    \\
    &=
    \hbar\mathbf{k}\cdot
    \bra{\mathbf{k},\bm\alpha}\hat{\mathbf{p}}(t)\ket{\mathbf{k},\bm\alpha},
    \label{eq:momAC:main}
\end{align}
where $\hat{\mathbf{p}}(t)=e^{i\hat{H}t/\hbar}\hat{\mathbf{p}}e^{-i\hat{H}t/\hbar}$ is the momentum operator in the Heisenberg picture.

Now, consider the thermal average of the chosen state $\ket{\bm\alpha}$ over thermal distribution $P(\bm\alpha)$ of coherent states \cite{glauber1963} where
\begin{align*}
    P(\bm\alpha)
    =
    \prod_{\mathbf{q}}\left[\frac{e^{-\abs{\alpha_{\mathbf{q}}}^2/N_{\mathbf{q}l}}}{\pi N_{\mathbf{q}l}}\right].
\end{align*}
Then, the thermal average of the inverse momentum relaxation time is
\begin{align}
    \ev{\frac{1}{\tau_{tr}}}_{\mathrm{th}}
    &=
    -\sum_{\mathbf{q}}
    \frac{\mathbf{k}\cdot\mathbf{q}}{\mathbf{k}^2}
    \frac{2\pi g_{\mathbf{q}l}^2}{\hbar}
    [
    N_{\mathbf{q}l}
    \delta(\varepsilon(\mathbf{k+q})-\varepsilon(\mathbf{k})-\hbar\omega_{\mathbf{q}l})
    \notag
    \\
    &\qquad+
    (N_{\mathbf{q}l}+1)
    \delta(\varepsilon(\mathbf{k+q})-\varepsilon(\mathbf{k})+\hbar\omega_{\mathbf{q}l})
    ]
    \label{eq:tautrth:main}
\end{align}

On the other hand, the transport scattering rate calculated in Fock state description is the sum of the transition rates of phonon annihilation $\Gamma_{\mathbf{k}\to \mathbf{k}+\mathbf{q}}^{(\mathrm{abs.})}$ and creation $\Gamma_{\mathbf{k}\to \mathbf{k}+\mathbf{q}}^{(\mathrm{emi.})}$ from initial state $\ket{\mathbf{k}}$ to any final state $\ket{\mathbf{k}\pm\mathbf{q}}$ and weighted by the geometric factor $-\frac{\mathbf{k}\cdot\mathbf{q}}{\mathbf{k}^2}$:
\begin{align}
    \ev{\frac{1}{\tau_{tr}}}_{\mathrm{th}}
    &=
    -\sum_{\mathbf{q}}
    \frac{\mathbf{k}\cdot\mathbf{q}}{\mathbf{k}^2}
    [\Gamma_{\mathbf{k}\to \mathbf{k}+\mathbf{q}}^{(\mathrm{abs.})}
    +
    \Gamma_{\mathbf{k}\to \mathbf{k}+\mathbf{q}}^{(\mathrm{emi.})}
    ]
    \label{invTauIEScatt:main}
\end{align}
where the transition rates of phonon annihilation and creation processes using Fermi's golden rule are
\begin{align*}
    \Gamma_{\mathbf{k}\to \mathbf{k}+\mathbf{q}}^{(\mathrm{abs.})}
    &=
    \frac{2\pi}{\hbar}
    g_{\mathbf{q}l}^2
    N_{\mathbf{q}l}\delta(\varepsilon(\mathbf{k}+\mathbf{q})-\varepsilon(\mathbf{k})-\hbar\omega_{\mathbf{q}l})
    \\
    \Gamma_{\mathbf{k}\to \mathbf{k}+\mathbf{q}}^{(\mathrm{emi.})}
    &=
    \frac{2\pi}{\hbar}
    g_{\mathbf{q}l}^2
    (N_{\mathbf{q}l}+1)\delta(\varepsilon(\mathbf{k}+\mathbf{q})-\varepsilon(\mathbf{k})+\hbar\omega_{\mathbf{q}l}),
\end{align*}
respectively.
It is clearly seen Eqs. \eqref{eq:tautrth:main}
and \eqref{invTauIEScatt:main} are exactly the same, showing the equivalence of coherent state and Fock state descriptions.

Considering Fermi statistics with Fermi--Dirac distribution $f(E)=(e^{\beta(E-\mu(T))}+1)^{-1}$ where $\mu(T)$ is the chemical potential and $\beta=1/k_BT$, Eq. \eqref{eq:tautrth:main} becomes
\begin{align}
    \ev{\frac{1}{\tau_{tr}}}
    &=
    -\beta
    \int_0^{\infty}\dd\varepsilon(\mathbf{k})
    \sum_{\mathbf{q}}
    \frac{\mathbf{k}\cdot\mathbf{q}}{\mathbf{k}^2}
    \frac{2\pi g_{\mathbf{q}l}^2}{\hbar}
    \notag
    \\
    &\times[
    N_{\mathbf{q}l}
    \delta(\varepsilon(\mathbf{k+q})-\varepsilon(\mathbf{k})-\hbar\omega_{\mathbf{q}l})
    \notag
    \\
    &\qquad+
    (N_{\mathbf{q}l}+1)
    \delta(\varepsilon(\mathbf{k+q})-\varepsilon(\mathbf{k})+\hbar\omega_{\mathbf{q}l})
    ]\notag
    \\
    &\times
    f(\varepsilon(\mathbf{k}))(1-f(\varepsilon(\mathbf{k}+\mathbf{q}))))
    \label{eq:considerfermi:main}
\end{align}
where each scattering process is weighted by the product of the probability $f(\varepsilon(\mathbf{k}))$ that the initial state of energy $\varepsilon(\mathbf{k})$ is occupied and the probability $1-f(\varepsilon(\mathbf{k}+\mathbf{q}))$ that the final state of energy $\varepsilon(\mathbf{k}+\mathbf{q})=\varepsilon(\mathbf{k})\pm\hbar\omega_{\mathbf{q}l}$ ($+$ and $-$ for a phonon absorption and emission, respectively) is unoccupied.

\begin{figure}[t]
\centering
\includegraphics[width=0.49\textwidth]{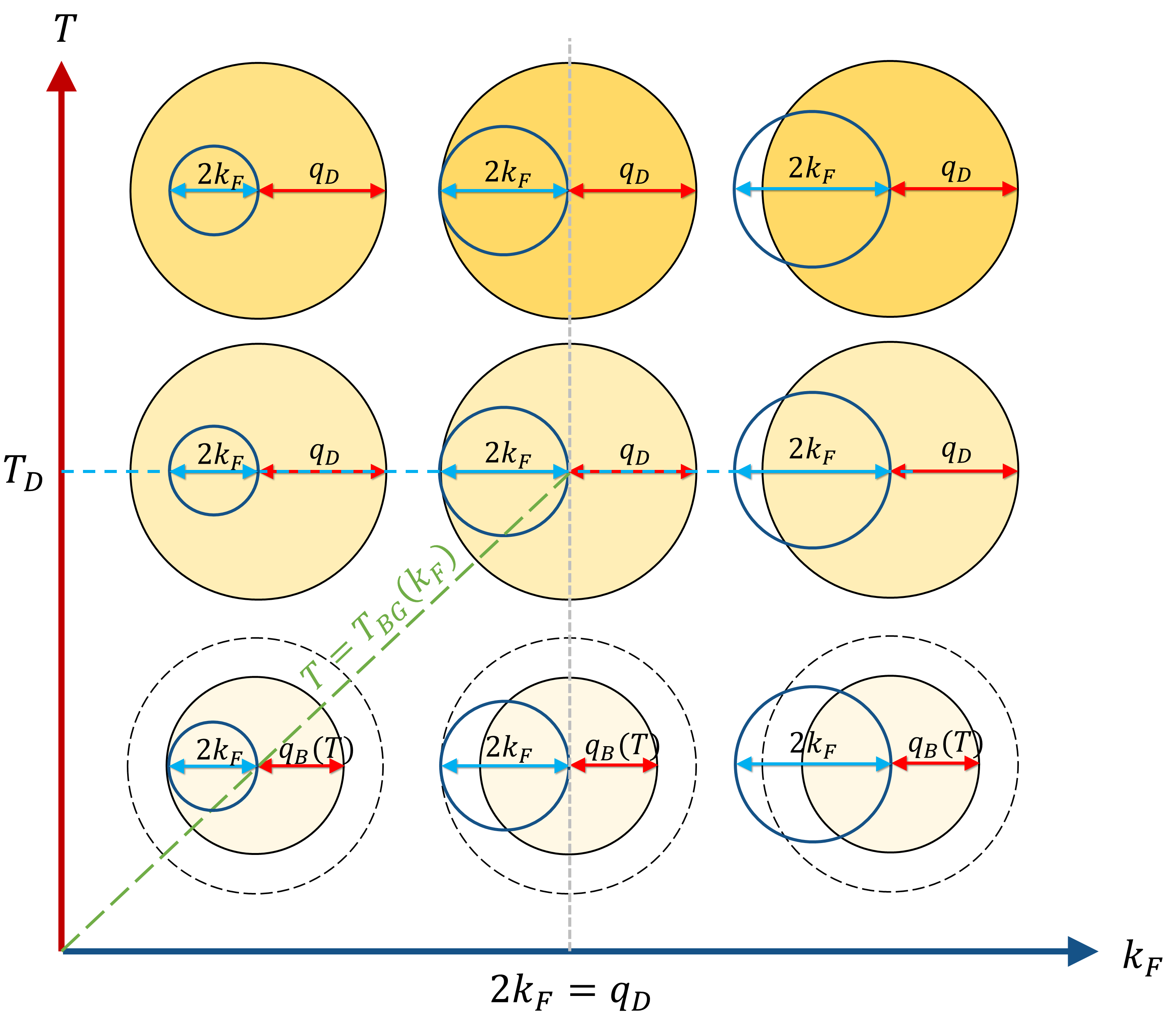}
\caption{Reciprocal space diagrams for given Fermi wave number $k_F$ and temperature $T$.
The contour of equal electronic energy is drawn as a blue circle with a radius $k_F$ varying only about the horizontal axis.
Fourier transform of the deformation potential is shown as a yellow disk depending only on temperature $T$, the vertical axis.
The radius of the yellow disk is given by the Bose wave number $q_B(T)$  up to which normal modes are thermally occupied. The color density of the disk reflects the occupation number of the lattice modes, which increases with temperature. A dashed black circle shows the Debye cutoff $q_D$.
The dashed green line shows the Bloch--Gr\"{u}neisen temperature $T_{\mathrm{BG}}(k_F)=\hbar v_s2k_F/k_B.$
Possible scattering processes appear at the intersections of the blue circle and yellow disk.}
\label{phasediareciprocal}
\end{figure}

The momentum relaxation process can be easily demonstrated with the reciprocal space diagram for the scattering shown in Fig. \ref{phasediareciprocal}.
The contour of equal electronic energy is drawn as a blue circle with a radius $k_F$ varying only about the horizontal axis.
Fourier transform of the deformation potential is shown as a yellow disk depending only on temperature $T$, the vertical axis.
The radius of the yellow disk is determined by the Bose wave number $q_B(T)$ roughly up to which normal modes are thermally occupied. Also, the color density of yellow disk reflects the occupation number of each mode increasing with temperature. In addition, a dashed black circle shows the Debye cutoff $q_D$.
Then, the possible scattering processes appear at the intersections of the blue circle and yellow disk.

The critical temperature separating low- and high-temperature behaviors are different in the following two regimes:

(i) $2k_F> q_D$ (e.g., typical metals). The maximal scattering wave number is given by Debye wave number $q_D$, and the critical temperature is given by the corresponding Debye temperature $T_D=\hbar v_s q_D/k_B$.

(ii) $2k_F< q_D$ (e.g., semimetals).
The maximal scattering wave number is given by twice the Fermi wave number $2k_F$ (backscattering), and the critical temperature is given by the corresponding Bloch-Gr{\"u}neisen temperature $T_{\mathrm{BG}}=\hbar v_s 2k_F/k_B$.
Although there exist shorter wavelength modes ($q>2k_F$), they do not contribute to scattering the electron as there is no energy-conserving transition for them. We say the electron is ``transparent'' to the shorter wavelength modes ($q>2k_F$)~\cite{kim2022bragg}.

We define the regime-independent parameters: the maximal scattering wave number $q_{\mathrm{max}}=\min\{q_D,2k_F\}$ and the corresponding critical temperature $T_c=\hbar v_s q_{\mathrm{max}}/k_B=\min\{T_D,T_{\mathrm{BG}}\}$.
The critical temperature divides low and high-temperature behaviors \cite{Fuhrer}:

(i) For low temperature limit $T\ll T_c$, the number of thermally activated modes rises as $q_B(T)\sim T$ in 2D [$(q_{B}(T))^2\sim T^2$ in 3D], the square of the matrix element increases as $\abs{g_{\mathbf{q}l}}^2\sim T$, and the geometric factor by $1-\cos\theta\sim T^2$, giving $1/\tau\sim T^4$ in 2D ($\sim T^5$ in 3D).

(ii) For high-temperature limit $T\gg T_c$,
only the square of the matrix element increases with temperature $\abs{g_{\mathbf{q}l}}^2\sim T$, so resistivity goes as $1/\tau\sim T$.



\subsection{Comparison of the quantum and classical fields}
In Sec. \ref{s:EquivCohFock}, we have obtained the inverse momentum relaxation time in the presence of quantum field $\hat{V}$ in Eq. \eqref{eq:considerfermi:main}. We now want to get the corresponding expression for a classical field.
The classical field is constructed from the quantum field in Eq. \eqref{eq:Vhat:main}
\begin{align}
    V
    =
    \mel{\alpha}{\hat{V}}{\alpha}
    =
    \sum_{\mathbf{q}}
    g_{\mathbf{q}l}
    (\alpha_{\mathbf{q}l}
    +
    \alpha_{-\mathbf{q}l}^\dag)
    e^{i\mathbf{q}\cdot\mathbf{r}},
    \label{eq:Vcl}
\end{align}
which basically amounts to replacing annihilation and creation operators $a$ and $a^\dag$ to complex numbers $\alpha$ and $\alpha^*$. As the complex numbers commutes, i.e., $[\alpha,\alpha^*]=0$, it does not capture quantum fluctuations (zero point motion of each normal mode). This is clearly seen by doing time-dependent perturbation theory as in Appendix \ref{app:PertCoh}:
\begin{align}
    \ev{\frac{1}{\tau_{tr}^{(C)}}}
    &=
    -\beta
    \int_0^{\infty}\dd\varepsilon(\mathbf{k})
    \sum_{\mathbf{q}}
    \frac{\mathbf{k}\cdot\mathbf{q}}{\mathbf{k}^2}
    \frac{2\pi g_{\mathbf{q}l}^2}{\hbar}
    \notag
    \\
    &\times[
    N_{\mathbf{q}l}
    \delta(\varepsilon(\mathbf{k+q})-\varepsilon(\mathbf{k})-\hbar\omega_{\mathbf{q}l})
    \notag
    \\
    &\qquad+
    N_{\mathbf{q}l}
    \delta(\varepsilon(\mathbf{k+q})-\varepsilon(\mathbf{k})+\hbar\omega_{\mathbf{q}l})
    ]\notag
    \\
    &\times
    f(\varepsilon(\mathbf{k}))(1-f(\varepsilon(\mathbf{k}+\mathbf{q})))),
    \label{eq:considerfermiC:main}
\end{align}
which clearly shows spontaneous emission is missing in the classical field (cf. the quantum field in Eq. \eqref{eq:considerfermi:main}).
This result is natural since the missing quantum fluctuations in the classical field is associated with   spontaneous emission.
Since spontaneous emission is absent in a classical field,   detailed balance is broken and an electron in the classical field will heat up as time goes by.

We can further consider a classical time-independent (frozen) field.
The time dependence of the field comes from the  evolution of the normal modes, i.e. $\alpha_{\mathbf{q}l}(t)=\alpha_{\mathbf{q}l}e^{-i\omega_{\mathbf{q}l}t}$.
If  time evolution is ignored, then $\alpha_{\mathbf{q}l}(t)=\alpha_{\mathbf{q}l},$ which is equivalent to setting the frequencies to zero $\omega_{\mathbf{q}l}=0$ (except for those appearing in $g_{\mathbf{q}l}$ and $N_{\mathbf{q}l}$).
Thus, the inverse momentum relaxation time is 
\begin{align}
    \ev{\frac{1}{\tau_{tr}^{(C_0)}}}
    &=
    -\beta
    \int_0^{\infty}\dd\varepsilon(\mathbf{k})
    \sum_{\mathbf{q}}
    \frac{\mathbf{k}\cdot\mathbf{q}}{\mathbf{k}^2}
    \frac{2\pi g_{\mathbf{q}l}^2}{\hbar}
    \notag
    \\
    &\times
    2N_{\mathbf{q}l}
    \delta(\varepsilon(\mathbf{k+q})-\varepsilon(\mathbf{k}))
    \notag
    \\
    &\times
    f(\varepsilon(\mathbf{k}))(1-f(\varepsilon(\mathbf{k}+\mathbf{q})))),
    \label{eq:considerfermiC0:main}
\end{align}
which shows there is no phonon absorption or emission [cf. the classical time-dependent field in Eq. \eqref{eq:considerfermiC:main}].
Necessarily, the scattering in the frozen field is elastic \cite{davies_physics_1998} unlike  in a time-dependent field.

The results from the different fields in Eqs. \eqref{eq:considerfermi:main}, \eqref{eq:considerfermiC:main}, and \eqref{eq:considerfermiC0:main} can be condensed, in the quasielastic approximation, to the following equation:
\begin{align}
    \ev{\frac{1}{\tau_{tr}^{(F)}}}
    &=
    -\sum_{\mathbf{q}}
    \frac{\mathbf{k}\cdot\mathbf{q}}{\mathbf{k}^2}
    \frac{2\pi g_{\mathbf{q}l}^2}{\hbar}
    2N_{\mathbf{q}l}J_{\mathbf{q}l}^{(F)}
    \notag
    \\
    &\qquad\qquad\times\delta(\varepsilon(\mathbf{k+q})-\varepsilon(\mathbf{k}))|_{\varepsilon(\mathbf{k})=E_F}
    \label{eq:tautrall:main}
\end{align}
where $F$ is an index indicating which field (quantum field $F=Q$, classical field $F=C$, and classical time-independent field $F=C_0$) is used and $J_{\mathbf{q}l}^{(F)}$ is a factor appearing in the corresponding field choice
\begin{align*}
    J_{\mathbf{q}l}^{(Q)}&=(N_{\mathbf{q}l}+1)\beta\hbar\omega_{\mathbf{q}l},
    \\
    J_{\mathbf{q}l}^{(C)}&=(N_{\mathbf{q}l}+1/2)\beta\hbar\omega_{\mathbf{q}l},
    \\
    J_{\mathbf{q}l}^{(C_0)}&=1.
\end{align*}
Note $J_{\mathbf{q}l}^{(Q)}>J_{\mathbf{q}l}^{(C)}>J_{\mathbf{q}l}^{(C_0)}=1$ for $\beta\hbar\omega_{\mathbf{q}l}>0$. In the high-temperature limit, where all the phonon energies are small compared to thermal energy, all the factors become   unity $\lim\limits_{\beta\hbar\omega_{\mathbf{q}l}\to0}J_{\mathbf{q}l}^{(Q)}=\lim\limits_{\beta\hbar\omega_{\mathbf{q}l}\to0}J_{\mathbf{q}l}^{(C)}=J_{\mathbf{q}l}^{(C_0)}=1$, so there is no difference between the choice of fields in the high-temperature limit, within   perturbation theory.

\begin{figure}
    \centering
    \includegraphics[width=.48\textwidth]{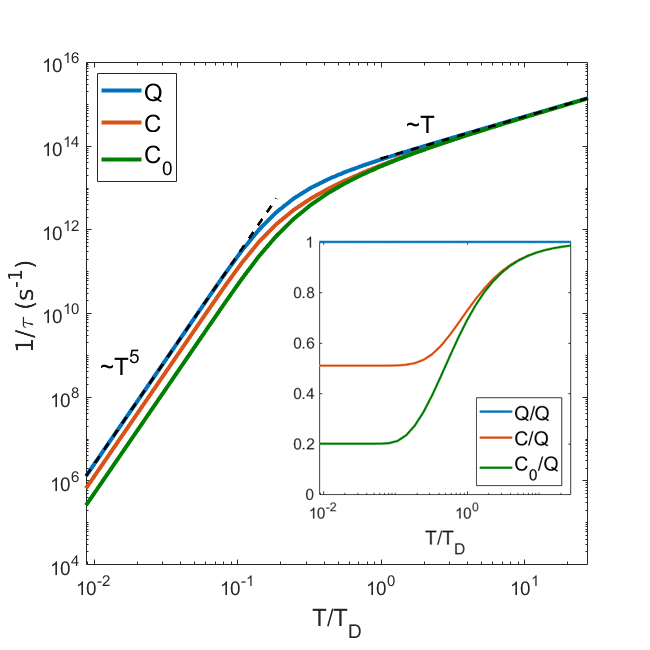}
    \caption{Temperature dependence of the inverse momentum relaxation time calculated using Eq. \eqref{invTau3DDP} for $2k_F/q_D=2.8$ in 3D. Blue, red, and green curves are from perturbation theory in quantum ($Q$), classical time-dependent ($C$), and classical time-independent ($C_0$) fields, respectively. The inset on the bottom right shows the ratios of the inverse momentum relaxation times to the one in the quantum field.
    The underestimation of the inverse momentum relaxation time in the classical fields is significant  near or below the critical temperature $T_c=\min\{T_D,T_{\mathrm{BG}}\}$ (in this case $T_c=T_D$).
    They all give correct temperature dependence in low- and high-temperature limits, but their proportionality constants in low-temperature limit differ as shown in the inset.}
    \label{InvTauFields}
\end{figure}

The underestimation of the inverse momentum relaxation time in the classical fields is significant  near or below the critical temperature $T_c=\min\{T_D,T_{\mathrm{BG}}\}$ as shown in Fig. \ref{InvTauFields}.
The inset of Fig. \ref{InvTauFields} shows the ratios of the inverse momentum relaxation times to the quantum field value.
The difference between the blue and red curves is from the missing spontaneous emission, which suppresses the rate by about a factor of 2.
The difference between the red and green curves arises from ignoring the time dependence of the potential (elastic approximation), which, with the missing spontaneous emission, decreases the rate by five (four) times in 3D (2D). The temperature dependence is, however, correctly captured with the classical fields,  allowing introduction of  a correction factor to reach the proper values, for example based on Fig. \ref{InvTauFields}.

\begin{figure}[t]
    \centering
    \includegraphics[width=.5\textwidth]{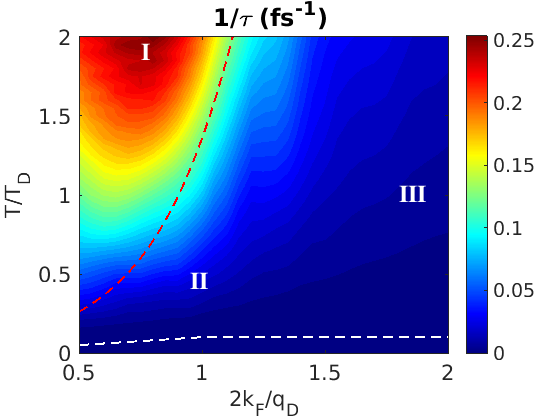}
    \caption{Temperature--Fermi momentum phase diagram from numerical results. 
    In region I, electrons are scattered by a strong deformation potential, causing electrons to localize under a frozen version of the  deformation potential.
  Region II shows perturbative and nonclassical behavior such as wave interference and diffraction.
    Region III shows perturbative and classical behavior.
The red dashed line ($V_{\textrm{rms}}/E_F=0.3$) divides perturbative and nonperturbative regions. 
The white dashed line ($T=0.1T_c$) divides the low- ($1/\tau\sim T^4$ for $T\ll T_c$) and high- ($1/\tau\sim T$ for $T\gg T_c$) temperature behaviors.}
    \label{fig:phasediaBoth}
\end{figure}

\section{Electron dynamics with coherent state lattice vibrations}\label{s:ElecDynamics}
In this section, we focus on nonperturbative electron dynamics under the coherent state description of lattice vibrations.
We use the split operator method (the codes can be found in Ref. \cite{MATLABcodePRB}) for electron wave-packet propagation~\cite{SplitOperator,Feit1982,Tannor}.
We launch  electron wave packets with an initial average momentum $\hbar k_F$ (to the $\hat{x}$ direction without loss of generality)
\begin{align*}
    \psi(x,y,0)=\frac{1}{\sqrt{2\pi\sigma_x\sigma_y}}\exp(-\frac{x^2}{4\sigma_x^2}-\frac{y^2}{4\sigma_y^2}+ik_Fx)
\end{align*}
on the deformation potential $V_D(\mathbf{r})$.
The initial wave function $\ket{\psi(0)}$ is propagated with the (split operator) propagator $U(t)$ to obtain the state $\ket{\psi(t)}=U(t)\ket{\psi(0)}$ at time $t$, and the average momentum $\ev{\hat{\mathbf{p}}}(t)=\mel{\psi(t)}{\hat{\mathbf{p}}}{\psi(t)}$ is calculated.

Within the relaxation time approximation, the magnitude of the average momentum is expected to decay as $\hbar k_Fe^{-t/\tau}$, which allows us to extract the inverse momentum relaxation time $1/\tau$ by fitting the average momentum to the exponential form.
For regions I and II in Fig. \ref{fig:phasediaBoth}, the average momentum shows non-exponential decay and the fit is not perfect. Nevertheless, the $1/\tau$ value still gives a qualitative description for the scattering rate.

Using this procedure, we construct a temperature--Fermi momentum phase diagram of the inverse momentum relaxation time $1/\tau$.
As in the reciprocal space diagram in Fig. \ref{phasediareciprocal}, we use dimensionless parameters $T/T_D$ and $2k_F/q_D$ for axes.
We have the following parameters  qualitatively affecting the  dynamics:

(i) $V_{\textrm{rms}}/E_F,$ the ratio of the potential fluctuation $V_{\textrm{rms}}$ to the average kinetic energy of the wave packet $E_F$.
$V_{\textrm{rms}}/E_F\ll1$ is where the potential is a perturbation to the free (effective mass) electron motion. This is the regime of  normal metals.
$V_{\textrm{rms}}/E_F\gtrsim1$ is the nonperturbative regime where the electron is strongly scattered, or even partially trapped by the potential, showing very different dynamics from the perturbed free (effective mass) electron motion.

(ii) $k_F/q_{\mathrm{eff}}(T)$ is the ratio of effective shortest wavelength $2\pi/q_{\mathrm{eff}}(T)$ of the potential to electron wavelength $2\pi/k_F$.
The classical regime is $k_F/q_{\mathrm{eff}}(T)\gg 1$ where the electron wavelength is  shorter than the effective shortest length scale of the deformation potential.

The temperature--Fermi momentum phase diagram (see typical metal parameters in Appendix \ref{app:paramnormal}) is shown in Fig.~\ref{fig:phasediaBoth}.
We divided the phase diagram into three different regions according to distinct behaviors they show.
There are no sharp boundaries between the regions; the changes are gradual, rather than phase transitions.
We explain  the distinct regions as follows:

(1) \textit{Region I}.
The highly nonperturbative ($V_{\textrm{rms}}/E_F>0.5$) localization region.
Electrons are scattered by a strong deformation potential, which would cause electrons to localize in short distances under a frozen deformation potential as a result of quantum interference effects. However, the time dependence of the potential will break this short-time localization, causing a transient localization~\cite{PhysRevB.75.235106,PhysRevB.83.081202,afmTL}.  This will be discussed in the next section.
The average momentum shows non-exponential decay due to quantum coherence and interference effects in a similar way as in Ref. \cite{kim2022bragg}.

\begin{figure}[t]
\centering
\includegraphics[width=0.23\textwidth]{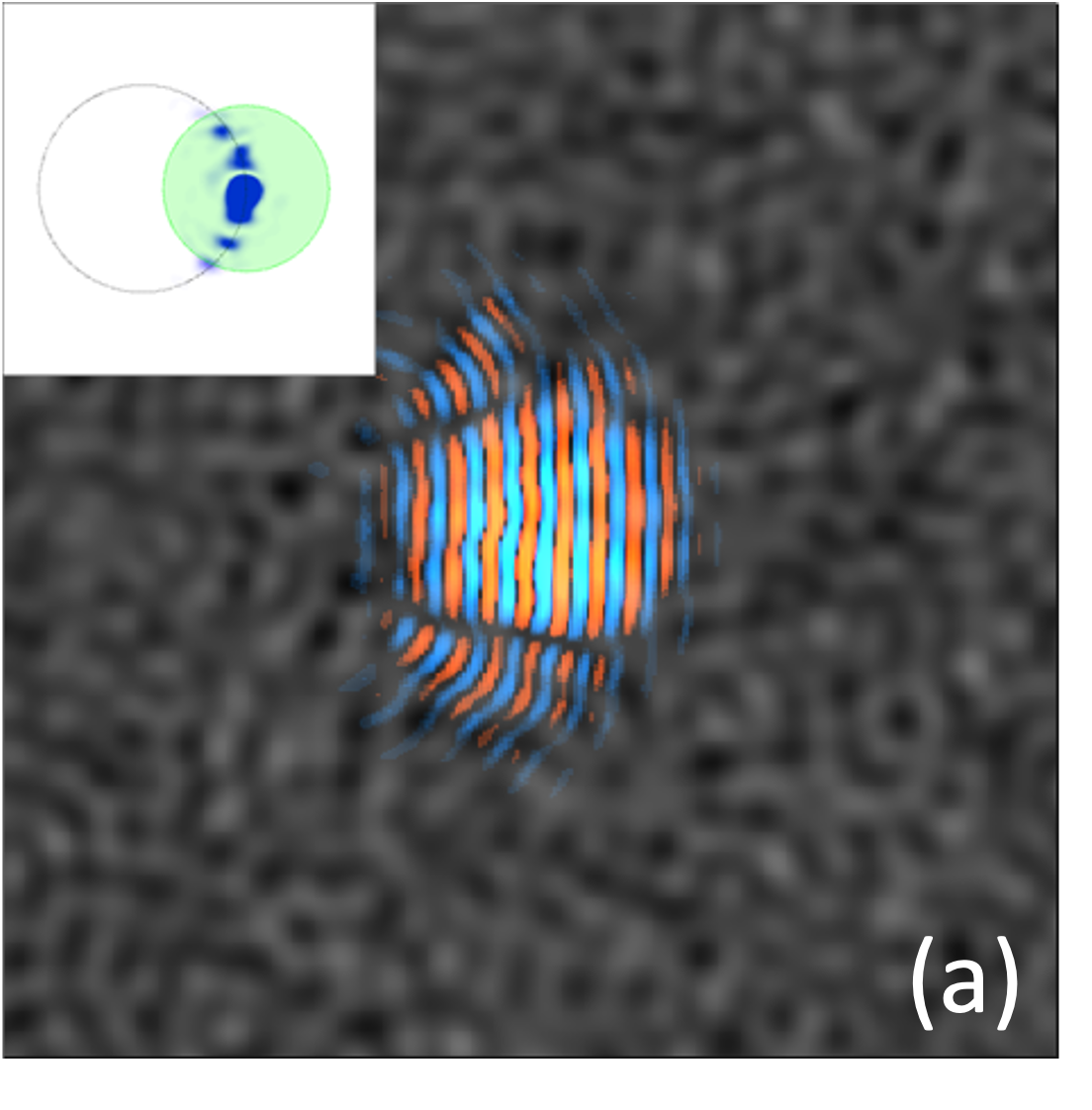}
\includegraphics[width=0.23\textwidth]{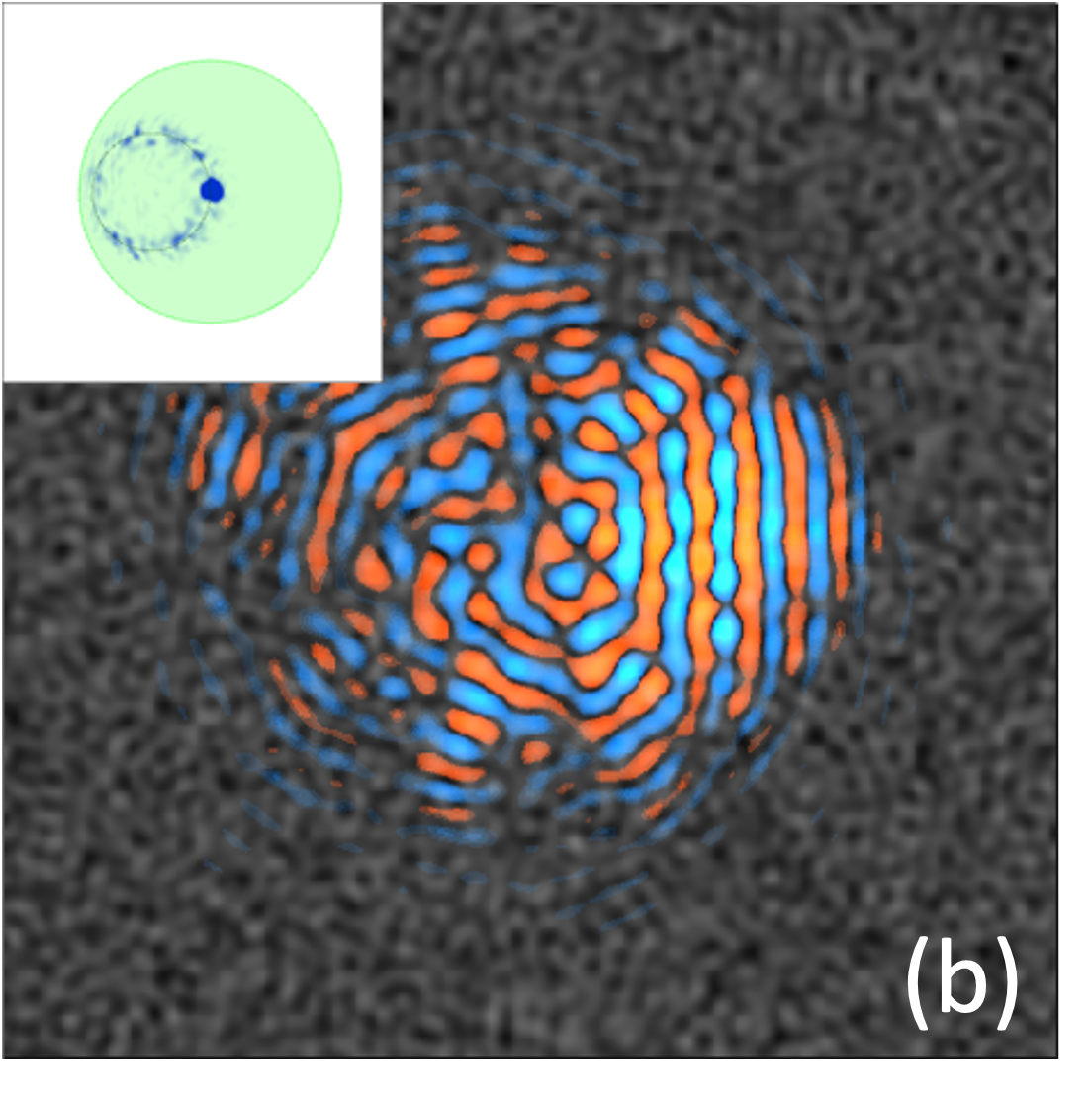}
\caption{A snapshot of a wave packet (shown in red/blue scale) initially launched rightward in the deformation potential (shown in gray scale) in (a) the classical regime (region III in Fig. \ref{fig:phasediaBoth}) and (b) the nonclassical regime (the left part of region II in Fig. \ref{fig:phasediaBoth}). The insets on the top left show the probability distributions of the wave in the momentum space. The constant electron energy contour (black dotted circle) is overlain by the nonzero Fourier components of the potential (light green disk). (a) Forward scattering is observed, which will lead to the branched flow at longer times \cite{branchedphysicstoday,kim2022bragg} (see Fig.~\ref{semicl}). (b) The scattering is diffractive and nearly isotropic.}
\label{fig:ClBeh}
\end{figure}

(2) \textit{Region II}.
This is the perturbative ($V_{\textrm{rms}}/E_F<0.15$) and nonclassical [$k_F\lesssim q_{\mathrm{eff}}(T)$] region.
Wave interference and diffraction effects are important because the electron wavelength is larger than the shortest length scale of the deformation potential as shown in Fig. \ref{fig:ClBeh}(b).
There is a partial transparency of the electrons to any shorter wavelength modes ($q>2k_F$) present in the underlying deformation potential~\cite{kim2022bragg} as was explained in Sec. \ref{s:EquivCohFock}.
Similar to region I, the average momentum shows non-exponential decay due to the quantum coherence and interference effects \cite{kim2022bragg}. 

\begin{figure}
    \centering
    \includegraphics[width=.45\textwidth]{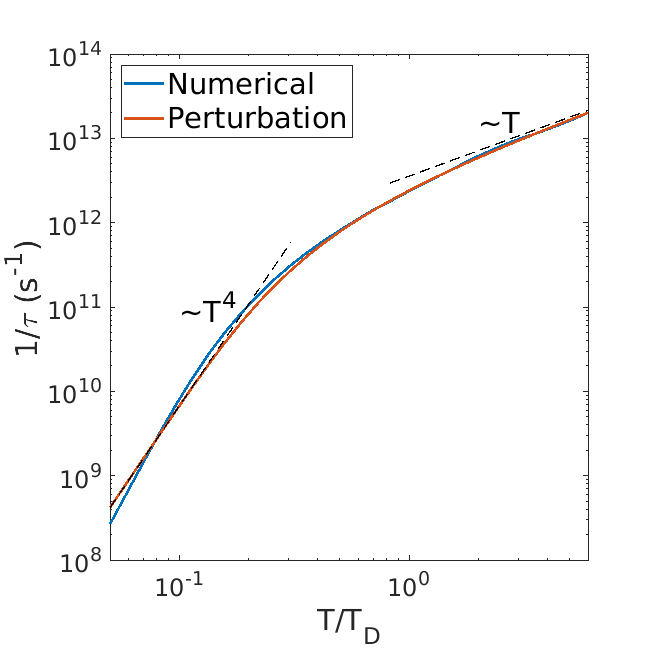}
    \caption{Temperature dependence of the inverse momentum relaxation time for $2k_F/q_D=2.8$. Blue curve is scattering rate obtained from exponential fitting of momentum decay in numerical results (vertical cross section of the phase diagram in Fig. \ref{fig:phasediaBoth} at $2k_F/q_D=2.8$). The red curve is taken from  perturbation theory [Eq. \eqref{invTau2DDP}] for $F=C_0$.}
    \label{TDep28}
\end{figure}

(3) \textit{Region III}.
This is the perturbative ($V_{\textrm{rms}}/E_F<0.15$) and classical [$k_F\gg q_{\mathrm{eff}}(T)$] regime.
Forward scattering is observed in Fig. \ref{fig:ClBeh}(a), which will lead to branched flow at longer times, such as shown in Fig. \ref{semicl}.
Branched flow regime is where propagating waves (or a collection of rays) form tree-like branches under a weakly disordered medium, due to small-angle refraction~\cite{branchedphysicstoday, daza2021propagation,kim2022bragg}.
Exponential decay of the average momentum of the wave packet is observed. In this  region  Fermi's golden rule works very well.
Comparison of the numerical result with the perturbation theory [Eq. \eqref{invTau2DDP} for $F=C_0$] is shown in Fig. \ref{TDep28}.
Blue and red curves compare numerical results with perturbation theory, respectively, showing a very good match.

We also ran  classical ray path simulations and checked that it gives consistent results with the quantum dynamics in region III (see more details in Appendix \ref{app:SemiclassicalRay}).
As the classical dynamics cannot capture wave nature of the electron such as interference and diffraction, it is not  valid in regions I and II.
Thus, in this work, we will focus on quantum dynamics in the main text, leaving the discussion of classical dynamics to Appendix \ref{app:SemiclassicalRay}.

\section{Charge carrier coherence effects and band tails}\label{s:coherence}

Coherent dynamics of a charge carrier interacting with lattice vibrations not only recovers the correct temperature dependence of resistivity in metals, but also leads to results which the conventional formalism (incoherent and uncorrelated succession of the first-order events through Boltzmann transport theory) is unable to capture. In this section, we discuss the charge carrier coherence effects that are carried beyond the single collision events, which may have important consequences for charge carrier transport.

\subsection{Transient localization at high-temperatures}
In our wave-packet simulations, we find charge carriers to be localized by the frozen deformation potential in long Fermi wavelength and high-temperature regime (region I in Fig. \ref{fig:phasediaBoth}). To investigate this regime further in detail, we choose parameters from a cuprate (see Appendix \ref{app:paramstrange}). 
In particular, we use $2k_F/q_D=0.5$, $T/T_D=1$, and $V_{\textrm{rms}}/E_F^{(h)}=1.87$ as our dimensionless parameters. Since the charge carriers in cuprates are holes, we use hole energy at the Fermi surface $E_F^{(h)}=E_{\mathrm{BM}}-E_F$ that is the difference between the band maximum energy $E_{\mathrm{BM}}$ and electron Fermi energy $E_F$. Holes are just the absence of electrons and thus discussions on electrons applies to holes as well.

\begin{figure}[t]
\centering
\includegraphics[width=0.48\textwidth]{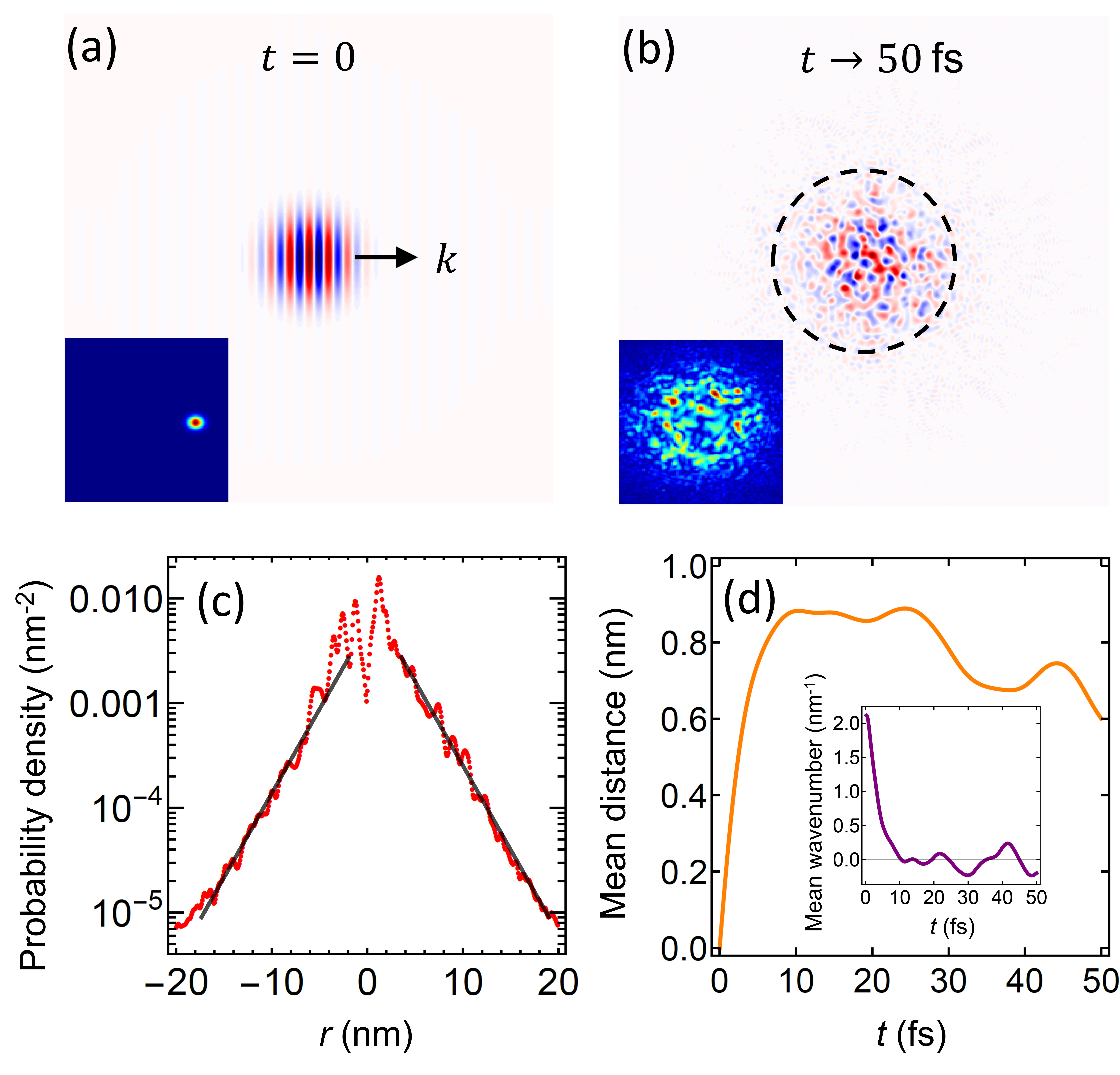}
\caption{Short-time localization of the wave packet at $2k_F/q_D=0.5$, $T/T_D=1$ and $V_{\textrm{rms}}/E_F^{(h)}=1.87$. (a) At $t=0$ the wave packet is launched to the right with an initial momentum $k$. On the bottom-left the momentum space picture is shown as inset. (b) Within a very short time $t<\tau_{\phi}$, the wave packet is localized by the frozen deformation potential. (c) Log-linear plot of radial probability density versus position (radial distance) indicates the exponential decay of the density profile radially in 2D space. (d) Fast saturation of the mean distance as well as a slight quantum-boomeranglike effect is shown. The inset shows the rapid decay of the average momentum to zero. After the rapid initial decay, average momentum of the wave packet oscillates around zero.}
\label{fig:localizationfig}
\end{figure}

The result of the simulation is shown in Fig.~\ref{fig:localizationfig}. The wave packet with an initial momentum $\hbar k_F$ [Fig.~\ref{fig:localizationfig}(a)] is launched toward the right in the frozen deformation potential. The momentum space picture is shown in the bottom-left inset. Figure \ref{fig:localizationfig}(b) shows a snapshot of the wave packet after $\SI{50}{fs}$ is passed. Due to the strong scattering and interference, the wave packet becomes localized to a zone in real space, indicated by the dashed black circle. 


We observe four different signatures of Anderson localization in this short-time behavior. (1) The exponential decay of radial probability density in Fig.~\ref{fig:localizationfig}(c). Linear wings in the log-linear plot suggests that the density profile decays exponentially. (2) The decrease of mean distance in the transport direction in Fig.~\ref{fig:localizationfig}(d), implying the wave packet tries to return toward its initial launch point, which is similar to the quantum boomerang effect~\cite{boomerang1}. (3) Rapid average momentum reversal in the inset of Fig.~\ref{fig:localizationfig}(d).  (4) The localized eigenstates obtained from the time Fourier transform of the wave function at given energies (see more details in Appendix \ref{app:localES}).

At time scales shorter than the characteristic timescale of the potential change, lattice vibrations act like static disorder. The short-time localization for the initial wave packet  happens around $\sim\SI{10}{fs}$ (Fig. \ref{fig:localizationfig}). This occurs earlier than the typical timescale for the potential change $\sim2\pi/\omega_D=\SI{125}{fs}$, so the frozen deformation potential is valid for the short-time dynamics. The other mechanisms for breaking electron coherence not considered here have similar or longer timescale. For a typical YBCO, electron-phonon inelastic scattering time is found to be $\tau_{e-ph}=\SI{100}{fs}$ at $\SI{400}{K}$~\cite{PhysRevLett.105.257001}. We also expect the electron dephasing time ($\tau_{\phi}$) to be in the order of $\tau_{e-ph}$~\cite{Lin_2002} whereas electron-electron scattering time  $\tau_{e-e}=\SI{1400}{fs}$ at $\SI{400}{K}$ which is much longer than other relevant time scales~\cite{PhysRevLett.105.257001}.

Nevertheless, what we observe here should not be interpreted as full Anderson localization of the charge carrier, since it only occurs under short-time dynamics. Because the deformation potential is actually time-dependent, at longer times, but before fully diffusive behavior is established, the lattice dynamics can delocalize the charge carriers, a phenomenon so-called transient localization~\cite{PhysRevB.75.235106,PhysRevB.83.081202,afmTL}.
In the transient localization, charge carriers encounter time varying landscapes of disorder, which breaks the quantum interference causing localization and initiates delocalization~\cite{PhysRevB.86.245201,PhysRevB.89.235201}. Due to strong scattering and transient localization effects, the charge carrier motion slows down. As a result, short-time localization as well as its breakdown at longer times may affect the charge transport properties in the materials showing bad/strange metal behavior. 



\subsection{Band tails in the density of states}

\begin{figure} 
    \centering
    \includegraphics[width=0.48\textwidth]{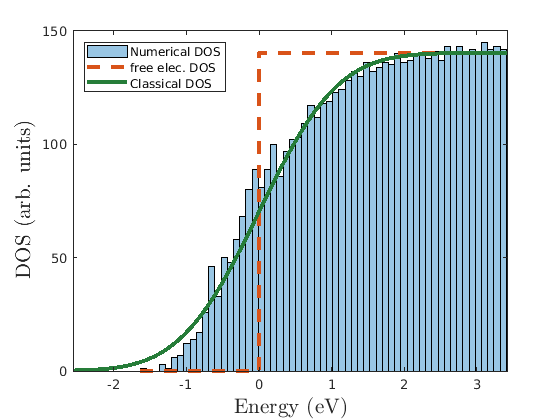}
    \caption{Band tail in a 2D density of states with a deformation potential present. $E=0$  is the threshold for the free electron without the deformation potential and the red dashed curve is its density of states. Blue bins are the histogram of energy eigenvalues obtained numerically and green curve is the classical density of states calculated from Eq. \eqref{eq:ClDOS}. The quantum density of states calculated numerically matches well with the classical density of states.}
    \label{fig:Lifschitz}
\end{figure}

The calculation of the quantum single electron density of states is simpler in the coherent state description.
By solving the time-independent Schr\"{o}dinger equation in Fourier space (basically using plane wave basis), we obtain eigenvalues and eigenstates, hence the quantum density of states as shown in Fig. \ref{fig:Lifschitz}. The quantum density of states obtained numerically is compared to the classical density of states given as
\begin{align}
    D_{\mathrm{Cl}}(E)=D_{\mathrm{free}}^{\mathrm{(2D)}}\frac{1+\erf(E/\sqrt{2}V_{\mathrm{rms}})}{2}
    \label{eq:ClDOS}
\end{align}
where $D_{\mathrm{free}}^{\mathrm{(2D)}}=\frac{m^*}{2\pi\hbar^2}$ is the free electron density of states in two dimensions, $\erf$ is the error function, and we did not include spin degeneracy.
Eq. \eqref{eq:ClDOS} is identical to the density of states expression in the presence of the high density of impurities with Gaussian statistics as shown in Ref. \cite{MieghemPhysRevB.44.12822} as the deformation potential also has Gaussian statistics (cf. Sec. \ref{s:DPproperties}).

\begin{figure}
    \centering
    \includegraphics[width=0.4\textwidth]{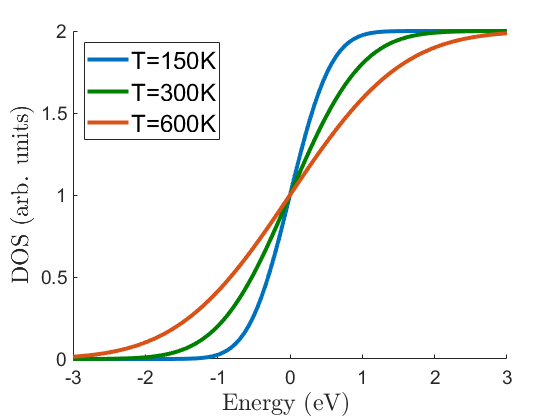}
    \caption{The temperature dependence of the band tails in the classical densities of states calculated from Eq. \eqref{eq:ClDOS} at three different temperatures. The widths of the tails are determined by the fluctuation of the potential $V_{\mathrm{rms}}(T)$. Unlike band tails from impurities, the tails from the deformation potential have temperature dependence.}
    \label{fig:DOST}
\end{figure}

The disorder in the deformation potential is, however, not from the actual impurities but rather from thermal fluctuations of the lattice which acts like an internal and slowly moving (with sound speed) impurity potential. Thus, the band tails from the deformation potential has temperature dependence where the tail width is determined by the fluctuation of the potential $V_{\mathrm{rms}}(T)$ as shown in Fig. \ref{fig:DOST}.

The theory of the band tails from the deformation potential, although it is not due to impurities, is relevant to large literature on the band tails in heavily doped semiconductors \cite{MieghemRevModPhys.64.755} including Halperin--Lax tails \cite{HalperinLaxPhysRev.148.722}, and the general
theory of disordered systems or localized
states in amorphous materials involving Lifschitz tails \cite{simon_lifschitz_1985}.
The band tail in Eq. \eqref{eq:ClDOS} is classified as a Gaussian tail \cite{MieghemRevModPhys.64.755,HalperinLaxPhysRev.148.722}.




\section{Discussion and Conclusion}\label{s:conclusion}
We have introduced a coherent state description of lattice vibrations and recast interactions with electrons, providing a quite different   paradigm for electron-phonon interactions from the conventional Fock state description.
In the coherent state description, the quasiclassical field of the deformation potential acts as an internal field on electrons, preserving their coherence for many collisions. 
This is analogous to an electron subject to a classical blackbody electromagnetic radiation field; photons are not counted, and the back-action of an electron on the incident field is neglected~\cite{milonni_introduction_2019}.
In contrast, in the Bloch-Gr\"uneisen theory within the Fock state description of lattice vibrations, any electron coherence effects lasting longer than a single collision time are lost, since higher-order scattering is approximated as an incoherent and uncorrelated concatenation of first-order scattering events through Boltzmann transport~\cite{ziman2001electrons,DasSarmaPRB2009}.

We successfully performed both semiclassical (in Appendix \ref{app:SemiclassicalRay}) and fully quantum (in the main text) analysis of the problem using ray trajectory and wave-packet propagation calculations, respectively.
We not only  find an agreement with Bloch-Gr\"uneisen theory in the weak field limit, but also go beyond by retaining the charge carrier coherence over longer times, revealing high-temperature coherence effects. We conclude that the charge carriers can remain  coherent in spite of conversing with huge numbers of phonons, in an analogy with  electrons in an electromagnetic field.

For strong fields at high-temperatures, the ``attempted" localization introduces modifications on carrier transport, which might be related to the exotic phases of strange metals. The violation of Mott--Ioffe--Regel limit and nonsaturation of resistivity of some metals at high-temperatures, for example, can be a consequence of the transient localization phenomenon, which hinders the transport~\cite{PhysRevB.75.235106,PhysRevB.83.081202,afmTL}.



In this work, we have considered the lattice to be  a smooth continuum. Nonetheless, it is also possible to embed the problem within a discrete model: time-dependent tight-binding calculation, including the effect of lattice vibrations as time-dependent hopping parameters, as was carried out for graphene within the time-dependent Schr\"odinger equation with no adiabatic Born--Oppenheimer approximation ~\cite{mohanty_lazy_2019}.

Here, our attention focuses on  charge carrier motion in a deformation field, ignoring  any back-action of the charge carrier on the field. Particularly, the local lattice response to a localized or quasi-trapped charge carrier has not been taken into account, but one can easily imagine polaronlike scenarios~\cite{Stoneham_2007, emin_2012, PhysRevB.54.12835, PhysRevB.32.3515, PhysRevB.14.3346, jacoboni1977review, PhysRevB.98.085201,PhysRevB.62.6317,PhysRevLett.81.2514,PhysRevLett.86.4624,PhysRevB.97.134305}.
Diagrammatic quantum Monte Carlo method treats the electron-optical phonon interactions nonperturbatively and gives useful insights on polaron problems \cite{PhysRevB.62.6317,PhysRevLett.81.2514,PhysRevLett.86.4624,PhysRevB.97.134305}.
The avenue is open to including back-action, respecting the fluctuation-dissipation theorem and electron-lattice vibrations coupling as a thermalization pathway.


In a recent paper~\cite{infrared}, Fratini and Ciuchi showed that ``slow boson modes'' explain the anomalous infrared displaced Drude peak behavior of the strange metals. They did not specify what these slow modes are. The deformation potential forms exactly the kind of ``slow moving bosonic field'' \cite{infrared}, or ``self-induced'' randomness~\cite{infrared2}, needed to explain the anomalous infrared Drude peak. The possible implications of our model in the case of displaced Drude peak in optical conductivity will be a future work.


The wave picture of the lattice vibrations introduced in this paper
opens vistas unavailable to the conventional particle picture implemented with perturbation theory.
We first show the consistency of the two pictures in certain regimes and then go beyond by revealing aspects that the conventional theory could not explain.
The nonperturbative treatment of the electron dynamics in the deformation potential has revealed the importance of electron coherence effects such as transient electron localization and band tails in the density of states. The coherent state paradigm of lattice vibrations gives insights on electron-phonon interactions and will serve as a theoretical framework to tackle important open questions in condensed matter physics.

The codes for the split operator method can be found in Ref. \cite{MATLABcodePRB}.

\begin{acknowledgments}
We thank Professor B. Halperin for many stimulating and informative discussions surrounding the issues raised in this paper. V. Mohanty helped inspire this work with his work on Ref. \cite{mohanty_lazy_2019}. Discussions with Professor S. Sachdev, Professor S. Das Sarma, Professor S. Kivelson, and Professor J. H. Miller Jr. were very helpful. We are very grateful to Professor P. Milonni for calling our attention to connections between  our work and the Hanbury Brown--Twiss literature and its aftermath.  We thank the National Science Foundation for supporting this research, through the NSF the Center for Integrated Quantum Materials (CIQM) Grant No. DMR-1231319. A.A. acknowledges support from The Scientific and Technological Research Council of Turkey (T\"urkiye Bilimsel ve Teknolojik Ara\c{s}t{\i}rma Kurumu, T\"UB\.{I}TAK) Program Code 2219. J.K.-R. thanks the Emil Aaltonen Foundation for financial support.
\end{acknowledgments}

\appendix
\section{MORE ON THE DEFORMATION POTENTIAL}
\subsection{General derivation: Lattices with noncubic (or nonsquare) symmetry}\label{app:GeneralDeriv}
In Sec. \ref{s:Deformation}, we regarded the lattice as having a cubic (or square in 2D) symmetry. However, this is not a serious restriction and can be generalized by starting with more general form of the expansion in Eq. \eqref{eq:DP2}:
\begin{align}
    E\left(\mathbf{k};\epsilon_{ij}(\mathbf{r})\right)
    &=
    E_0(\mathbf{k})
    +
    \sum_{i,j}E_d^{ij}(\mathbf{k})\epsilon_{ij}(\mathbf{r})
    +
    \cdots\label{eq:DP2Gen}
\end{align}
where $\mathbf{k}$ is an electron wave vector, $E_0(\mathbf{k})$ is band energy of undistorted (hence periodic) lattice, and $E_d^{ij}(\mathbf{k})$ are the expansion coefficients of the first-order terms.
Here, the first-order correction terms in the strain fields, $\sum_{i,j}E_d^{ij}(\mathbf{k})\epsilon_{ij}(\mathbf{r})$,
can be considered as the deformation energy.
The expansion coefficients $E_d^{ij}(\mathbf{k})$ have $\mathbf{k}$-dependence, but they can be treated as constants to a good approximation, particularly for nonpolar semiconductors~\cite{BS2}, and then $E_d^{ij}$ are called deformation potential constants. Thus, the deformation energy without the $\mathbf{k}$-dependence
\begin{align}
    V_D(\mathbf{r})=\sum_{i,j}E_d^{ij}\epsilon_{ij}(\mathbf{r})
    \label{eq:DPdefgen}
\end{align}
is defined as \textit{the deformation potential}. Equation \eqref{eq:DPdefgen} is the generalization of Eq. \eqref{eq:DPdef} in Sec. \ref{s:Deformation}; it does not assume cubic (or square) symmetry.
We can reduce Eq. \eqref{eq:DPdefgen} back to Eq. \eqref{eq:DPdef} by considering a lattice with cubic (or square) symmetry where $E_d^{ij}=0$ for $i\neq j$ and $E_d^{ii}=E_d$:
\begin{align}
    V_D(\mathbf{r})=E_d\nabla\cdot\mathbf{u}(\mathbf{r}),
\end{align}
where $\sum_i\epsilon_{ii}(\mathbf{r})=\nabla\cdot\mathbf{u}(\mathbf{r})$ is dilation.

\subsection{Classical derivation}\label{app:ClassicalDeriv}
From Eq. (A.28) of Bardeen and Shockley~\cite{BS2}, with a bit of generalization, the classical displacement field from a normal mode with a wave vector $\mathbf{q}$ and a polarization index $\lambda$ can be written as
\begin{align}
    \mathbf{u}_{\mathbf{q}\lambda}(\mathbf{x},t)
    =
    \bm{\varepsilon}_{\mathbf{q}\lambda}
    (
    A_{\mathbf{q}\lambda}e^{i\mathbf{q}\cdot\mathbf{x}-i\omega_{\mathbf{q}\lambda}t}
    +
    A_{\mathbf{q}\lambda}^* e^{-i\mathbf{q}\cdot\mathbf{x}+i\omega_{\mathbf{q}\lambda}t}
    ),
    \label{ClDispl}
\end{align}
where $\bm{\varepsilon}_{\mathbf{q}\lambda}$, $\omega_{\mathbf{q}\lambda}$, and $A_{\mathbf{q}\lambda}$ are polarization unit vector, angular frequency, and complex amplitude, respectively, of a normal mode indexed by $\mathbf{q}\lambda$.
Note $\mathbf{x}$ is a spatial coordinate that has nothing to do with electron coordinate $\mathbf{r}$ so far.
The total displacement field is the superposition of the displacement fields from the normal modes:
\begin{align*}
    \mathbf{u}(\mathbf{x},t)
    =
    \sum_{\mathbf{q}\lambda}
    \mathbf{u}_{\mathbf{q}\lambda}(\mathbf{x},t).
\end{align*}
The Hamiltonian for a normal mode $\mathbf{q}\lambda$ is given by
\begin{align}
    H_{\mathbf{q}\lambda}
    &=
    \int\dd\mathbf{x}
    \left[
    \frac{1}{2}\rho_m\left(\pdv{\mathbf{u}_{\mathbf{q}\lambda}}{t}\right)^2
    +
    \frac{1}{2}\rho_m\omega_{\mathbf{q}\lambda}^2\mathbf{u}_{\mathbf{q}\lambda}^2
    \right]
    \label{ClHam}
    \\
    &=
    2\rho_m \mathcal{V}\omega_{\mathbf{q}\lambda}^2\abs{A_{\mathbf{q}\lambda}}^2,
    \notag
\end{align}
where $\rho_m$ and $\mathcal{V}$ are mass density and volume of the solid, respectively.
For a system in thermal equilibrium, the average Hamiltonian for each normal mode is given by
\begin{align*}
    \ev{H_{\mathbf{q}\lambda}}_{\textrm{th}}=N_{\mathbf{q}\lambda}\hbar\omega_{\mathbf{q}\lambda},
\end{align*}
where $N_{\mathbf{q}\lambda}=1/(e^{\hbar\omega_{\mathbf{q}\lambda}/k_BT}-1)$ is Bose occupation.
Taking the thermal average value, we obtain
\begin{align*}
    \abs{A_{\mathbf{q}\lambda}}
    =
    \sqrt{\frac{N_{\mathbf{q}\lambda}\hbar}{2\rho_m \mathcal{V}\omega_{\mathbf{q}\lambda}}}.
\end{align*}
Also, the phase $\varphi_{\mathbf{q}\lambda}=\arg(A_{\mathbf{q}\lambda})$ of a normal mode $\mathbf{q}\lambda$ can be defined such that
$A_{\mathbf{q}\lambda}
    =
    \abs{A_{\mathbf{q}\lambda}}e^{i\varphi_{\mathbf{q}\lambda}}$.

In the deformation potential model, the effective potential for the electron at a position $\mathbf{r}$ is determined by the strain field exactly at the same position $\mathbf{r}$ due to the local approximation.
Thus, the classical deformation potential field can be written as
\begin{align*}
    V_D(\mathbf{r},t)
    &=E_d\nabla\cdot\mathbf{u}(\mathbf{r},t)
    \\
    &=
    E_d
    \sum_{\mathbf{q}}
    iq
    (
    A_{\mathbf{q}l}e^{i\mathbf{q}\cdot\mathbf{r}-i\omega_{\mathbf{q}l}t}
    -
    A_{\mathbf{q}l}^* e^{-i\mathbf{q}\cdot\mathbf{r}+i\omega_{\mathbf{q}l}t}
    ),
\end{align*}
where $E_d$ is the deformation potential constant and $l$ stands for longitudinal acoustic mode. Note only longitudinal acoustic modes $(\mathbf{q}\cdot\bm{\varepsilon}_{\mathbf{q}l}=q)$ have a contribution to the deformation potential, not transverse modes $(\mathbf{q}\cdot\bm{\varepsilon}_{\mathbf{q}t}=0)$, as the transverse modes do not change the interatomic spacing up to the first-order in lattice distortion. 
We use the Debye model that introduces linear dispersion $\omega_{\mathbf{q}l}=v_s|\mathbf{q}|$ where $v_s$ is sound speed and Debye wave number (isotropic cutoff) $q_D$. Thus, we obtain the classical deformation potential field
\begin{align*}
    V_D(\mathbf{r},t)
    &=
    -E_d
    \sum_{\substack{\mathbf{q}\\q<q_D}}\sqrt{\frac{2N_{\mathbf{q}l}\hbar}{\rho_m \mathcal{V}\omega_{\mathbf{q}l}}}
    q
    \sin(\mathbf{q}\cdot\mathbf{r}-\omega_{\mathbf{q}l}t+\varphi_{\mathbf{q}l}),
\end{align*}
where $\omega_{\mathbf{q}l}=v_sq$.
Trivially, the phases $\varphi_{\mathbf{q}l}$ can be redefined to get another equivalent form of the formula, \textit{e.g.}, the one with cosine instead of sine.


\subsection{spatiotemporal autocorrelation function}\label{app:autocorr}
The spatiotemporal autocorrelation function of the deformation potential gives the strength of the potential fluctuation and the decay of its spatiotemporal correlation:
\begin{align}
    C(\bm{\delta}\mathbf{r},\delta t)
    &=
    \ev{V_D(\mathbf{r},t)V_D(\mathbf{r}+\bm{\delta}\mathbf{r},t+\delta t)}\notag
    \\
    &=
    (\mathcal{V}\mathcal{T})^{-1}
    \int_\mathcal{V}\dd\mathbf{r}
    \int_0^{\mathcal{T}}\dd t
    V_D(\mathbf{r},t)V_D(\mathbf{r}+\bm{\delta}\mathbf{r},t+\delta t)\notag
    \\
    &=
    \sum_{\substack{\mathbf{q}\\\abs{\mathbf{q}}<q_D}}
    4g_{\mathbf{q}l}^2 N_{\mathbf{q}l}
    \cos(\mathbf{q}\cdot\bm{\delta}\mathbf{r}-\omega_{\mathbf{q}}\delta t)/2\notag
    \\
    &=
    \int_{\abs{\mathbf{q}}<q_D}\frac{\mathcal{V}\dd\mathbf{q}}{(2\pi)^d}
    4g_{\mathbf{q}l}^2 N_{\mathbf{q}l}
    \cos(\mathbf{q}\cdot\bm{\delta}\mathbf{r}-\omega_{\mathbf{q}}\delta t)/2\label{eq:ACFinal}
\end{align}
where $d$ is the dimension of the considered system.
The autocorrelation is significant for spatiotemporal relation $\delta r=v_s\delta t$ corresponding to the sound wave propagation.
From the autocorrelation, we can obtain the typical energy scale of the potential fluctuation, i.e., the root mean square of the potential values
\begin{align*}
    V_{\textrm{rms}}=\sqrt{\ev{(V_D(\mathbf{r},t))^2}}=\sqrt{C(0,0)}.
\end{align*}
Note although the electron-phonon coupling strength of each mode $g_{\mathbf{q}l}\sim1/\sqrt{\mathcal{V}}$ has a volume dependence, the potential fluctuation $V_{\textrm{rms}}$ does not. This is because the number of modes $\sim\mathcal{V}$ cancel the volume dependence out as shown in $\mathcal{V}g_{\mathbf{q}l}^2$ factor in Eq. \eqref{eq:ACFinal}.

For one dimension $d=1$,
\begin{align*}
    C^{\mathrm{(1D)}}(\delta r,\delta t)
    &=
    \int_0^{q_D}
    \frac{\dd q}{2\pi}
    \frac{E_d^22\hbar q}{\rho_m v_s}
    \frac{\cos(q\delta r-v_sq\delta t)}
    {e^{\hbar v_s q/k_BT}-1}.
\end{align*}

\begin{figure}
    \centering
    \includegraphics[width=.35\textwidth]{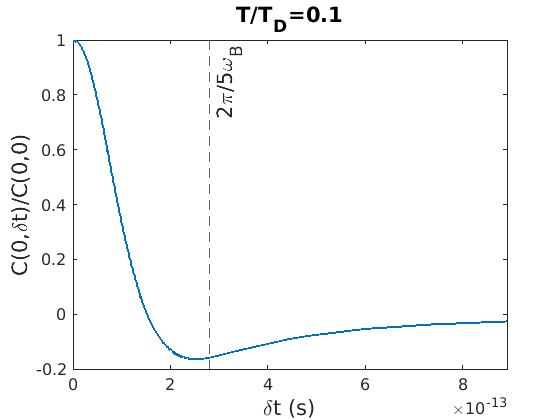}
    \includegraphics[width=.35\textwidth]{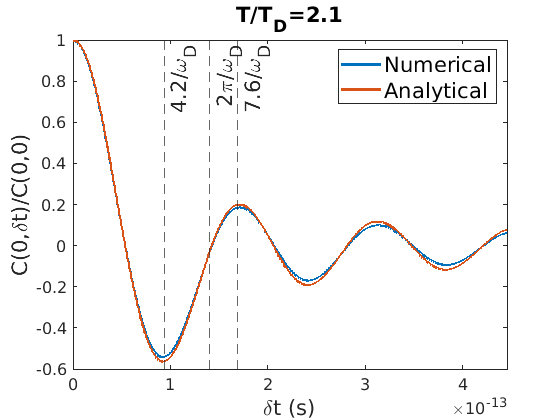}
    \caption{Temporal autocorrelation functions at the two different temperatures $T=0.1T_D$ (Top panel) and $T=2.1T_D$ (Bottom panel).
    (top panel) The blue curve is from the evaluation of the analytical expression \eqref{eq:AC2D}. (bottom panel) The blue curve (``Numerical'') is from the numerical evaluation of the autocorrelation $C(0,\delta t)=\ev{V_D(\mathbf{r},t)V_D(\mathbf{r},t+\delta t)}$ for one realization of the deformation potential, and the red curve (``Analytical'') is from the evaluation of the analytical expression \eqref{eq:AC2D}.
    At $T=0.1T_D$, $2\pi/5\omega_B$ is the timescale of the autocorrelation decay.
    At $T=2T_D$, the autocorrelation has first negative and positive peaks at $4.2/q_D$ and $7.6/q_D$, respectively, both are in a similar order as the characteristic timescale of the cutoff $2\pi/\omega_D$.}
    \label{ACDPT2D}
\end{figure}

For two dimensions $d=2$,
\begin{align*}
    C^{\mathrm{(2D)}}(\delta r,\delta t)
    &=
    \int_0^{q_D}
    \frac{\dd qq}{(2\pi)^2}
    \frac{E_d^22\hbar q}{\rho_m v_s}
    \frac{\pi J_0(q\delta r)}
    {e^{\hbar v_s q/k_BT}-1}
    \cos(v_sq\delta t)
\end{align*}
where we used
\begin{align*}
    \int_0^\pi\dd\theta \cos(A\cos\theta)
    =
    \pi J_0(\abs{A})
\end{align*}
Spatial autocorrelation function $C^{\mathrm{(2D)}}(\delta r,0)$ is shown in Fig. \ref{ACDP}. Temporal autocorrelation function $C^{\mathrm{(2D)}}(0,\delta t)$ is shown in Fig. \ref{ACDPT2D}.

For three dimensions $d=3$,
\begin{align*}
    C^{\mathrm{(3D)}}(\delta r,\delta t)
    &=
    \int_0^{q_D}
    \frac{\dd qq^2}{(2\pi)^3}
    \frac{E_d^22\hbar q}{\rho_m v_s}
    \frac{2\pi\mathrm{sinc}(q\delta r)}
    {e^{\hbar v_s q/k_BT}-1}
    \cos(v_sq\delta t)
\end{align*}
where $\mathrm{sinc}(x)=\sin x/x$ for $x\neq0$, $\mathrm{sinc}(0)=1$.

\subsection{Fourier transform of the potential}
The Fourier transform of the deformation potential is
\begin{align*}
    \tilde{V}_D(\mathbf{q},t)
    &=
    \frac{1}{\mathcal{V}}
    \int\dd\mathbf{r}
    e^{-i\mathbf{q}\cdot\mathbf{r}}
    V_D(\mathbf{r},t)
    =
    \mel{\mathbf{k}+\mathbf{q}}{V_D(t)}{\mathbf{k}}
    \\
    &=
    g_{\mathbf{q}l}\sqrt{N_{\mathbf{q}l}}
    (e^{i(-\omega_{\mathbf{q} l}t + \varphi_{\mathbf{q}})}
    +
    e^{-i(-\omega_{\mathbf{q} l}t + \varphi_{-\mathbf{q}})}).
\end{align*}
The absolute square of this quantity is
\begin{align}
    \abs{\tilde{V}_D(\mathbf{q},t)}^2
    &=
    2g_{\mathbf{q}l}^2 N_{\mathbf{q}l}
    \Theta(q_D-q),
    \label{DefPotMatEl}
\end{align}
where $\Theta$ is a unit step function.

\subsection{Observation of localized eigenstates}\label{app:localES}
\begin{figure}
    \centering
    \includegraphics[width=.35\textwidth]{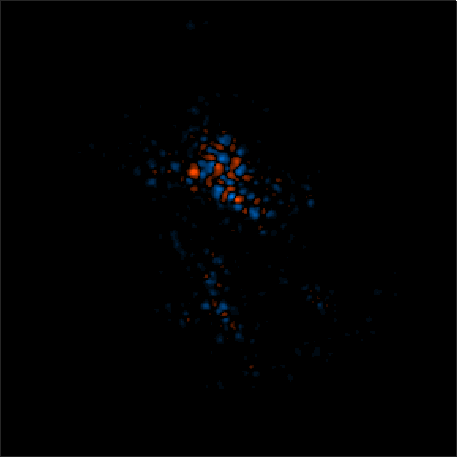}
    \caption{The eigenstate obtained from Eq. \eqref{eq:ES} for energy eigenvalue $E=0.545V_{\textrm{rms}}$ for the deformation potential at $T/T_D=10$.}
    \label{ALES}
\end{figure}

To obtain an  eigenstate $\phi_\mathcal{T}(E)$ of energy $E$ within the time window $\mathcal{T}$, one should calculate the time Fourier transform of the wave function~\cite{Heller:way}.
\begin{align}
    \ket{\phi_\mathcal{T}(E)}=\frac{1}{\pi\hbar}\int_0^\mathcal{T}\dd te^{iEt/\hbar}\ket{\psi(t)}.
    \label{eq:ES}
\end{align}
Also, the time Fourier transform of the autocorrelation function gives spectrum.
\begin{align}
    S_\mathcal{T}(E)=\frac{1}{\pi\hbar}\int_0^\mathcal{T}\dd te^{iEt/\hbar}\bra{\psi(0)}\ket{\psi(t)},
    \label{eq:Spec}
\end{align}
where $S_\mathcal{T}(E)$ is normalized such that $\int_{-\infty}^{\infty}\dd E S_\mathcal{T}(E)=1$.
Figure \ref{ALES} shows a localized eigenstate of energy $E=0.545V_{\textrm{rms}}$.


\section{COMPARISON OF COHERENT AND FOCK STATE PICTURES IN PERTURBATION THEORY}\label{app:PertCohFock}
\subsection{Perturbation theory in coherent state picture}\label{app:PertCoh}
Consider a system of a harmonic solid with electrons and quantized lattice vibrations. For simplicity, we think of a single-electron Hamiltonian and contemplate Fermi statistics afterwards. Then, the system Hamiltonian is written as
\begin{align*}
    \hat{H}=\hat{H}_0+\hat{V},
\end{align*}
where
\begin{align*}
    \hat{H}_0
    =
    \frac{\hat{\mathbf{p}}^2}{2m^*}
    +
    \sum_{\mathbf{q}}\hbar\omega_{\mathbf{q}l}(a_{\mathbf{q}l}^\dag a_{\mathbf{q}l}+1/2)
\end{align*}
is the sum of the electronic kinetic energy and the elastic energy of lattice vibrations and 
\begin{align}
    \hat{V}
    =
    \sum_{\mathbf{q}}
    g_{\mathbf{q}l}
    (a_{\mathbf{q}l}
    +
    a_{-\mathbf{q}l}^\dag)
    e^{i\mathbf{q}\cdot\mathbf{r}},
    \label{eq:Vhat}
\end{align}
is their interaction energy in the Schr\"{o}dinger picture [cf. Eq. \eqref{VDquantumfield}].

Let us describe the lattice with a multimode coherent state $\ket{\bm\alpha}$ defined in Eq. \eqref{eq:Vecalpha}.
Then, consider a scattering of an initial many-body state $\ket{\mathbf{k},\bm\alpha}$, where $\mathbf{k}$ is an electron wave vector, by the quantum deformation field $\hat{V}$.
The time-dependent perturbation theory is used by treating $\hat{H}_0$ as the unperturbed Hamiltonian and $\hat{V}$ as a perturbation \cite{sakurai2011modern}.
Then, in the interaction picture,
\begin{align*}
    \hat{V}_I(t)&=e^{i\hat{H}_0t/\hbar}\hat{V}e^{-i\hat{H}_0t/\hbar},
    \\
    \hat{\mathbf{p}}_I(t)
    &=
    e^{i\hat{\mathbf{p}}^2t/2m^*\hbar}\hat{\mathbf{p}}e^{-i\hat{\mathbf{p}}^2t/2m^*\hbar}
    =
    \hat{\mathbf{p}}.
\end{align*}
The inverse of the momentum relaxation time can be calculated from the momentum autocorrelation
\begin{align}
    C(t)
    &=
    \bra{\mathbf{k},\bm\alpha}\frac{\hat{\mathbf{p}}\cdot\hat{\mathbf{p}}(t)+\hat{\mathbf{p}}(t)\cdot\hat{\mathbf{p}}}{2}\ket{\mathbf{k},\bm\alpha}
    \\
    &=
    \hbar\mathbf{k}\cdot
    \bra{\mathbf{k},\bm\alpha}\hat{\mathbf{p}}(t)\ket{\mathbf{k},\bm\alpha},
    \label{eq:momAC}
\end{align}
where $\hat{\mathbf{p}}(t)=e^{i\hat{H}t/\hbar}\hat{\mathbf{p}}e^{-i\hat{H}t/\hbar}$ is the momentum operator in the Heisenberg picture. The average momentum in the interaction picture can be written as

\begin{widetext}

\begin{align}
    \bra{\mathbf{k},\bm\alpha}\hat{\mathbf{p}}(t)\ket{\mathbf{k},\bm\alpha}
    &=
    \prescript{}{I}{\bra{\mathbf{k},\bm\alpha(t)}}\hat{\mathbf{p}}_I(t)\ket{\mathbf{k},\bm\alpha(t)}_I
    \\
    &=
    \bra{\mathbf{k},\bm\alpha}\hat{\mathbf{p}}_I(t)\ket{\mathbf{k},\bm\alpha}
    +
    \prescript{}{I}{\mel{\mathbf{k},\bm\alpha^{(1)}(t)}
    {\hat{\mathbf{p}}_I(t)}
    {\mathbf{k},\bm\alpha}}
    +
    \mel{\mathbf{k},\bm\alpha}
    {\hat{\mathbf{p}}_I(t)}
    {\mathbf{k},\bm\alpha^{(1)}(t)}_I
    \\
    &\quad+
    \prescript{}{I}{\bra{\mathbf{k},\bm\alpha^{(1)}(t)}}
    \hat{\mathbf{p}}_I(t)
    \ket{\mathbf{k},\bm\alpha^{(1)}(t)}_I
    +
    \prescript{}{I}{\mel{\mathbf{k},\bm\alpha^{(2)}(t)}
    {\hat{\mathbf{p}}_I(t)}
    {\mathbf{k},\bm\alpha}}
    +
    \mel{\mathbf{k},\bm\alpha}
    {\hat{\mathbf{p}}_I(t)}
    {\mathbf{k},\bm\alpha^{(2)}(t)}_I
    +
    \cdots,
    \label{eq:avgmom}
\end{align}
\end{widetext}
where the terms were expanded up to the second-order in the perturbation and
\begin{align*}
    \ket{\mathbf{k},\bm\alpha^{(1)}(t)}_I
    &=
    \int_0^t\frac{\dd t'}{i\hbar}\hat{V}_I(t')\ket{\mathbf{k},\bm\alpha},
    \\
    \ket{\mathbf{k},\bm\alpha^{(2)}(t)}_I
    &=
    \int_0^t\frac{\dd t'}{i\hbar}\int_0^{t'}\frac{\dd t''}{i\hbar}\hat{V}_I(t')\hat{V}_I(t'')\ket{\mathbf{k},\bm\alpha}
\end{align*}
are the first- and second-order corrections to the wave function. Note the time derivatives of the terms appearing in the average momentum Eq. \eqref{eq:avgmom} are vanishing up to the first-order:
\begin{align*}
    \dv{\bra{\mathbf{k},\bm\alpha}\hat{\mathbf{p}}_I(t)\ket{\mathbf{k},\bm\alpha}}{t}
    &=
    \dv{(\hbar\mathbf{k})}{t}
    =0,
    \\
    \dv{\mel{\mathbf{k},\bm\alpha}
    {\hat{\mathbf{p}}_I(t)}
    {\mathbf{k},\bm\alpha^{(1)}(t)}_I}{t}
    +
    \textrm{c.c.}
    &=
    \frac{\mathbf{k}}{i}\bra{\mathbf{k},\bm\alpha}
    \hat{V}_I(t)\ket{\mathbf{k},\bm\alpha}
    +
    \textrm{c.c.}
    \\
    &=
    0
\end{align*}
and non-vanishing in the second order:
\begin{widetext}
\begin{align*}
    \dv{\bra{\mathbf{k},\bm\alpha^{(1)}(t)}
    \hat{\mathbf{p}}_I(t)
    \ket{\mathbf{k},\bm\alpha^{(1)}(t)}}{t}
    &=
    \int_0^t\frac{\dd t'}{\hbar^2}\bra{\mathbf{k},\bm\alpha}\hat{V}_I(t)
    \hat{\mathbf{p}}_I(t)
    \hat{V}_I(t')\ket{\mathbf{k},\bm\alpha}
    +\text{c.c.}
    \\
    &=
    \int_0^t\frac{\dd t'}{\hbar^2}
    \sum_{\mathbf{q}}
    \hbar(\mathbf{k}+\mathbf{q})
    \bra{\mathbf{k},\bm\alpha}\hat{V}_I(t)
    \ket{\mathbf{k}+\mathbf{q}}\bra{\mathbf{k}+\mathbf{q}}
    \hat{V}_I(t')\ket{\mathbf{k},\bm\alpha}
    +\text{c.c.},
    \\
    \dv{\mel{\mathbf{k},\bm\alpha}
    {\hat{\mathbf{p}}_I(t)}
    {\mathbf{k},\bm\alpha^{(2)}(t)}}{t}
    +\text{c.c.}
    &=
    -\int_0^t\frac{\dd t'}{\hbar^2}\hbar\mathbf{k}\bra{\mathbf{k},\bm\alpha}\hat{V}_I(t)\hat{V}_I(t')\ket{\mathbf{k},\bm\alpha}
    +\text{c.c.}
    \\
    &=
    -\int_0^t\frac{\dd t'}{\hbar^2}
    \sum_{\mathbf{q}}
    \hbar\mathbf{k}\bra{\mathbf{k},\bm\alpha}\hat{V}_I(t)\ket{\mathbf{k}+\mathbf{q}}\bra{\mathbf{k}+\mathbf{q}}\hat{V}_I(t')\ket{\mathbf{k},\bm\alpha}
    +\text{c.c.}
\end{align*}
Thus, substituting the result into Eq. \eqref{eq:momAC}, we obtain the inverse of the momentum relaxation time (or transport lifetime)
\begin{align}
    \frac{1}{\tau_{tr}}
    &=
    -\dv{[C(t)/C(0)]}{t}
    =
    -\int_0^t\frac{\dd t'}{\hbar^2}
    \sum_{\mathbf{q}}
    \frac{\mathbf{k}\cdot\mathbf{q}}{\mathbf{k}^2}
    \bra{\mathbf{k},\bm\alpha}\hat{V}_I(t)
    \ket{\mathbf{k}+\mathbf{q}}\bra{\mathbf{k}+\mathbf{q}}
    \hat{V}_I(t')\ket{\mathbf{k},\bm\alpha}
    +\text{c.c.},
    \label{eq:momrelax}
\end{align}
where
\begin{align}
    &\quad\bra{\mathbf{k},\bm\alpha}\hat{V}_I(t)
    \ket{\mathbf{k}+\mathbf{q}}\bra{\mathbf{k}+\mathbf{q}}
    \hat{V}_I(t')\ket{\mathbf{k},\bm\alpha}
    \\
    &=
    g_{\mathbf{q}l}^2
    e^{-i[\varepsilon(\mathbf{k+q})-\varepsilon(\mathbf{k})](t-t')/\hbar}
    \bra{\bm\alpha}
    (a_{\mathbf{q}l}^\dag e^{i\omega_{\mathbf{q}l}t}
    +
    a_{-\mathbf{q}l} e^{-i\omega_{\mathbf{q}l}t})
    (a_{\mathbf{q}l}e^{-i\omega_{\mathbf{q}l}t'}
    +
    a_{-\mathbf{q}l}^\dag e^{i\omega_{\mathbf{q}l}t'})
    \ket{\bm\alpha}.
    \label{eq:alphaAvg}
\end{align}
Now, consider the thermal average of the chosen state $\ket{\bm\alpha}$ over thermal distribution $P(\bm\alpha)$ of coherent states \cite{glauber1963} where
\begin{align*}
    P(\bm\alpha)
    =
    \prod_{\mathbf{q}}\left[\frac{e^{-\abs{\alpha_{\mathbf{q}}}^2/N_{\mathbf{q}l}}}{\pi N_{\mathbf{q}l}}\right].
\end{align*}
Then, the thermal average of the factor in Eq. \eqref{eq:alphaAvg} is
\begin{align*}
    &\quad\int\dd\bm\alpha P(\bm\alpha)
    \bra{\bm\alpha}
    (a_{\mathbf{q}l}^\dag e^{i\omega_{\mathbf{q}l}t}
    +
    a_{-\mathbf{q}l} e^{-i\omega_{\mathbf{q}l}t})
    (a_{\mathbf{q}l}e^{-i\omega_{\mathbf{q}l}t'}
    +
    a_{-\mathbf{q}l}^\dag e^{i\omega_{\mathbf{q}l}t'})
    \ket{\bm\alpha}
    \\
    &=
    \int\dd\bm\alpha P(\bm\alpha)
    \left[
    \abs{\alpha_{\mathbf{q}l}}^2 e^{i\omega_{\mathbf{q}l}(t-t')}
    +
    (\abs{\alpha_{-\mathbf{q}l}}^2+1)
    e^{-i\omega_{\mathbf{q}l}(t-t')}
    \right]
    \\
    &=
    N_{\mathbf{q}l} e^{i\omega_{\mathbf{q}l}(t-t')}
    +
    (N_{\mathbf{q}l}+1)
    e^{-i\omega_{\mathbf{q}l}(t-t')}.
\end{align*}
Thus, the thermal average of the inverse momentum relaxation time in Eq. \eqref{eq:momrelax} is
\begin{align*}
    \ev{\frac{1}{\tau_{tr}}}_{\mathrm{th}}
    =
    -\sum_{\mathbf{q}}
    \frac{\mathbf{k}\cdot\mathbf{q}}{\mathbf{k}^2}
    \frac{2g_{\mathbf{q}l}^2}{\hbar^2}
    \left[
    N_{\mathbf{q}l}
    \frac{\sin((\varepsilon(\mathbf{k+q})-\varepsilon(\mathbf{k})-\hbar\omega_{\mathbf{q}l})t/\hbar)}{(\varepsilon(\mathbf{k+q})-\varepsilon(\mathbf{k})-\hbar\omega_{\mathbf{q}l})/\hbar}
    +
    (N_{\mathbf{q}l}+1)
    \frac{\sin((\varepsilon(\mathbf{k+q})-\varepsilon(\mathbf{k})+\hbar\omega_{\mathbf{q}l})t/\hbar)}{(\varepsilon(\mathbf{k+q})-\varepsilon(\mathbf{k})+\hbar\omega_{\mathbf{q}l})/\hbar}
    \right].
\end{align*}
When the time $t$ is sufficiently large, the summand can be well approximated as delta functions
\begin{align}
    \ev{\frac{1}{\tau_{tr}}}_{\mathrm{th}}
    =
    -\sum_{\mathbf{q}}
    \frac{\mathbf{k}\cdot\mathbf{q}}{\mathbf{k}^2}
    \frac{2\pi g_{\mathbf{q}l}^2}{\hbar}
    [
    N_{\mathbf{q}l}
    \delta(\varepsilon(\mathbf{k+q})-\varepsilon(\mathbf{k})-\hbar\omega_{\mathbf{q}l})
    +
    (N_{\mathbf{q}l}+1)
    \delta(\varepsilon(\mathbf{k+q})-\varepsilon(\mathbf{k})+\hbar\omega_{\mathbf{q}l})
    ]
    \label{eq:tautrth}.
\end{align}
\end{widetext}

\subsection{Perturbation theory in Fock state picture}\label{app:PertFock}

The derivation in Appendix \ref{app:PertCoh} is equally applicable to Fock states. Consider a scattering of an initial many-body eigenstate $\ket{\mathbf{k},\mathbf{n}}$, labeled by the electron wave vector $\mathbf{k}$ and a collection of phonon occupation numbers $\mathbf{n}=(\dots,n_{\mathbf{q}\lambda},\dots)$ from normal modes $\mathbf{q}\lambda$'s, by the perturbation $\hat{V}$ defined in Eq. \eqref{eq:Vhat}.
The time-dependent perturbation theory for the eigenstate $\ket{\mathbf{k},\mathbf{n}}$ of the unperturbed Hamiltonian $\hat{H}_0$ reduces to Fermi's golden rule.
Then, the inverse of the momentum relaxation time (or transport lifetime) for the initial state $\ket{\mathbf{k},\mathbf{n}}$ is determined by the sum of the transition rates $\Gamma_{\mathbf{k},\mathbf{n}\to \mathbf{k}+\mathbf{q},\mathbf{n}'}$ from initial state $\ket{\mathbf{k},\mathbf{n}}$ to any final state $\ket{\mathbf{k}+\mathbf{q},\mathbf{n}'}$ and weighted by the factor $-\frac{\mathbf{k}\cdot\mathbf{q}}{\mathbf{k}^2}$ [cf. Eq. \eqref{eq:momrelax}]: 
\begin{align}
    \frac{1}{\tau_{tr}}
    &=
    -\sum_{\mathbf{q},\mathbf{n}'}
    \frac{\mathbf{k}\cdot\mathbf{q}}{\mathbf{k}^2}
    \Gamma_{\mathbf{k},\mathbf{n}\to \mathbf{k}+\mathbf{q},\mathbf{n}'},
    \label{invTauIE}
\end{align}
where the transition rates are given by Fermi's golden rule
\begin{align*}
    \Gamma_{\mathbf{k},\mathbf{n}\to \mathbf{k}+\mathbf{q},\mathbf{n}'}
    &=
    \frac{2\pi}{\hbar}
    \abs{\mel{\mathbf{k}+\mathbf{q},\mathbf{n}'}{\hat{V}}{\mathbf{k},\mathbf{n}}}^2
    \\
    &\qquad\times\delta(\varepsilon(\mathbf{k}+\mathbf{q},\mathbf{n}')-\varepsilon(\mathbf{k},\mathbf{n}))
\end{align*}
and
\begin{align*}
    \varepsilon(\mathbf{k},\mathbf{n})
    =
    \frac{\hbar^2\mathbf{k}^2}{2m^*}
    +
    \sum_{\mathbf{q}}\hbar\omega_{\mathbf{q}l}(n_{\mathbf{q}l}+1/2)
\end{align*}
is many-body eigenvalue for $\ket{\mathbf{k},\mathbf{n}}$.
Then, the Eq. \eqref{invTauIE} can be written as
\begin{align}
    \frac{1}{\tau_{tr}}
    &=
    -\sum_{\mathbf{q}}
    \frac{\mathbf{k}\cdot\mathbf{q}}{\mathbf{k}^2}
    (\Gamma_{\mathbf{k}\to \mathbf{k}+\mathbf{q}}^{(\mathrm{abs.})}
    +
    \Gamma_{\mathbf{k}\to \mathbf{k}+\mathbf{q}}^{(\mathrm{emi.})}),
    \label{invTauIEScatt}
\end{align}
where
\begin{align*}
    \Gamma_{\mathbf{k}\to \mathbf{k}+\mathbf{q}}^{(\mathrm{abs.})}
    &=
    \frac{2\pi}{\hbar}
    g_{\mathbf{q}l}^2
    n_{\mathbf{q}l}\delta(\varepsilon(\mathbf{k}+\mathbf{q})-\varepsilon(\mathbf{k})-\hbar\omega_{\mathbf{q}l})
    \\
    \Gamma_{\mathbf{k}\to \mathbf{k}+\mathbf{q}}^{(\mathrm{emi.})}
    &=
    \frac{2\pi}{\hbar}
    g_{\mathbf{q}l}^2
    (n_{\mathbf{q}l}+1)\delta(\varepsilon(\mathbf{k}+\mathbf{q})-\varepsilon(\mathbf{k})+\hbar\omega_{\mathbf{q}l})
\end{align*}
are scattering rates associated with phonon absorption and emission, respectively.

Now, consider the thermal average of the chosen state $\ket{\mathbf{n}}$ over thermal distribution $P_F(\mathbf{n})=\prod_{\mathbf{q}}\left[e^{-\beta \hbar\omega_{\mathbf{q}l}(n_{\mathbf{q}l}+1/2)}\right]$. 
Then, the thermal average of the inverse momentum relaxation time in Eq. \eqref{invTauIEScatt} is exactly identical to Eq. \eqref{eq:tautrth}, showing the equivalence of coherent state and Fock state descriptions.

\subsection{Considering Fermi statistics}
Considering Fermi statistics, Eq. \eqref{eq:tautrth} becomes
\begin{widetext}
\begin{align}
    \ev{\frac{1}{\tau_{tr}}}
    &=
    -\beta
    \int_0^{\infty}\dd\varepsilon(\mathbf{k})
    \sum_{\mathbf{q}}
    \frac{\mathbf{k}\cdot\mathbf{q}}{\mathbf{k}^2}
    \frac{2\pi g_{\mathbf{q}l}^2}{\hbar}
    \bigr[
    N_{\mathbf{q}l}
    \delta[\varepsilon(\mathbf{k+q})-\varepsilon(\mathbf{k})-\hbar\omega_{\mathbf{q}l}]
    f(\varepsilon(\mathbf{k}))\{1-f[\varepsilon(\mathbf{k})+\hbar\omega_{\mathbf{q}l}]\}
    \notag
    \\
    &\quad+
    (N_{\mathbf{q}l}+1)
    \delta[\varepsilon(\mathbf{k+q})-\varepsilon(\mathbf{k})+\hbar\omega_{\mathbf{q}l}]
    f(\varepsilon(\mathbf{k}))\{1-f[\varepsilon(\mathbf{k})-\hbar\omega_{\mathbf{q}l}]\}
    \bigr],
    \label{eq:considerfermi}
\end{align}
where each scattering process is weighted by the product of the probability $f(\varepsilon(\mathbf{k}))$ that the initial state of energy $\varepsilon(\mathbf{k})$ is occupied and the probability $1-f(\varepsilon(\mathbf{k}+\mathbf{q}))$ that the final state of energy $\varepsilon(\mathbf{k}+\mathbf{q})=\varepsilon(\mathbf{k})\pm\hbar\omega_{\mathbf{q}l}$ ($+$ and $-$ for a phonon absorption and emission, respectively) is unoccupied.

In the quasielastic approximation, phonon energies are far smaller than the electronic energies so that the electronic energy remains almost the same after creating or annihilating a phonon $\hbar\omega_{\mathbf{q}l}\ll\varepsilon(\mathbf{k}+\mathbf{q}),\varepsilon(\mathbf{k})$. Then, the factor $f(\varepsilon(\mathbf{k}))(1-f(\varepsilon(\mathbf{k})\pm\hbar\omega_{\mathbf{q}l}))$ is narrowly peaked around $\varepsilon(\mathbf{k})\approx\mu(T)$ with a characteristic width $k_BT$, for the temperature $k_BT\ll E_F$.
Note also $\mu(T)\approx E_F$ for $k_BT\ll E_F$. 
Thus, the remaining integrand can be expanded at $\varepsilon(\mathbf{k})=E_F$ (or $k=k_F$) and taking the lowest order (zeroth-order) term gives
\begin{align}
    \ev{\frac{1}{\tau_{tr}}}
    &=
    -\sum_{\mathbf{q}}
    \frac{\mathbf{k}\cdot\mathbf{q}}{\mathbf{k}^2}
    \frac{2\pi g_{\mathbf{q}l}^2}{\hbar}
    [
    N_{\mathbf{q}l}
    \delta(\varepsilon(\mathbf{k+q})-\varepsilon(\mathbf{k})-\hbar\omega_{\mathbf{q}l})
    g(\hbar\omega_{\mathbf{q}l})
    +
    (N_{\mathbf{q}l}+1)
    \delta(\varepsilon(\mathbf{k+q})-\varepsilon(\mathbf{k})+\hbar\omega_{\mathbf{q}l})
    g(-\hbar\omega_{\mathbf{q}l})
    ]|_{\varepsilon(\mathbf{k})=E_F}
    \label{invTauIEScattInteg}
    \\
    &\approx
    -\sum_{\mathbf{q}}
    \frac{\mathbf{k}\cdot\mathbf{q}}{\mathbf{k}^2}
    \frac{2\pi g_{\mathbf{q}l}^2}{\hbar}
    2N_{\mathbf{q}l}(N_{\mathbf{q}l}+1)\beta\hbar\omega_{\mathbf{q}l}
    \delta(\varepsilon(\mathbf{k}+\mathbf{q})-\varepsilon(\mathbf{k}))|_{\varepsilon(\mathbf{k})=E_F},
    \label{invTauIEScattInteg2}
\end{align}
\end{widetext}
where we defined a function
\begin{align*}
    g(u)
    &
    =
    \int_{0}^{\infty}\dd(\beta\varepsilon)f(\varepsilon)(1-f(\varepsilon+u))
    \\
    &\approx
    \int_{-\infty}^{\infty}\dd(\beta\varepsilon)f(\varepsilon)(1-f(\varepsilon+u))
    =
    \frac{e^{\beta u}\beta u}{e^{\beta u}-1}
\end{align*}
and used $g(\hbar\omega_{\mathbf{q}l})=(N_{\mathbf{q}l}+1)\beta \hbar\omega_{\mathbf{q}l}$ and $g(-\hbar\omega_{\mathbf{q}l})=N_{\mathbf{q}l}\beta \hbar\omega_{\mathbf{q}l}$.
Note considering Fermi statistics gave additional $g(\pm\hbar\omega_{\mathbf{q}l})$ factor in Eq. \eqref{invTauIEScattInteg} compared to Eq. \eqref{eq:tautrth}.

\subsection{Considering classical time-dependent and time-independent fields}
In Appendix \ref{app:PertCoh}, 2, 3, we have obtained the inverse momentum relaxation time in the presence of quantum field $\hat{V}$ in Eq. \eqref{invTauIEScattInteg2}. We now want to get the corresponding expression for a classical field.
The classical field is constructed from the quantum field in Eq. \eqref{eq:Vhat}
\begin{align}
    V
    =
    \mel{\bm\alpha}{\hat{V}}{\bm\alpha}
    =
    \sum_{\mathbf{q}}
    g_{\mathbf{q}l}
    (\alpha_{\mathbf{q}l}
    +
    \alpha_{-\mathbf{q}l}^*)
    e^{i\mathbf{q}\cdot\mathbf{r}},
    \label{eq:Vcl:main}
\end{align}
which basically amounts to replacing annihilation and creation operators $a$ and $a^\dag$ to complex numbers $\alpha$ and $\alpha^*$. As the complex numbers commutes, i.e., $[\alpha,\alpha^*]=0$, it does not capture quantum fluctuations (zero point motion of each normal mode). This is clearly seen by doing time-dependent perturbation theory as in Appendix \ref{app:PertCoh}.
In the interaction picture the classical field is written as
\begin{align*}
    V_I=e^{i\hat{\mathbf{p}}^2t/2m^*\hbar}Ve^{-i\hat{\mathbf{p}}^2t/2m^*\hbar}.
\end{align*}
Fermi statistics can be considered as in Eq. \eqref{eq:considerfermi}. In the quasielastic approximation [cf. Eq. \eqref{invTauIEScattInteg}], the inverse momentum relaxation time in the classical time-dependent field is
\begin{widetext}
\begin{align}
    \ev{\frac{1}{\tau_{tr}^{(C)}}}
    &=
    -\sum_{\mathbf{q}}
    \frac{\mathbf{k}\cdot\mathbf{q}}{\mathbf{k}^2}
    \frac{2\pi g_{\mathbf{q}l}^2}{\hbar}
    [
    N_{\mathbf{q}l}
    \delta(\varepsilon(\mathbf{k+q})-\varepsilon(\mathbf{k})-\hbar\omega_{\mathbf{q}l})
    g(\hbar\omega_{\mathbf{q}l})
    +
    N_{\mathbf{q}l}
    \delta(\varepsilon(\mathbf{k+q})-\varepsilon(\mathbf{k})+\hbar\omega_{\mathbf{q}l})
    g(-\hbar\omega_{\mathbf{q}l})
    ]|_{\varepsilon(\mathbf{k})=E_F}
    \label{invTauCF}
    \\
    &\approx
    -\sum_{\mathbf{q}}
    \frac{\mathbf{k}\cdot\mathbf{q}}{\mathbf{k}^2}
    \frac{2\pi g_{\mathbf{q}l}^2}{\hbar}
    N_{\mathbf{q}l}
    (2N_{\mathbf{q}l}+1)\beta\hbar\omega_{\mathbf{q}l}\delta(\varepsilon(\mathbf{k+q})-\varepsilon(\mathbf{k}))|_{\varepsilon(\mathbf{k})=E_F}
    \label{eq:tautrthCF}
\end{align}
which clearly shows spontaneous emission is missing in the classical field [cf. the quantum field in Eq. \eqref{invTauIEScattInteg}].
This result is natural since the missing quantum fluctuations in the classical field are associated with the spontaneous emission.
Since spontaneous emission is absent in a classical field, detailed balance is broken and an electron in the classical field will heat up as time goes by.

We can further consider a classical time-independent (frozen) field.
The time dependence of the field comes from the  evolution of the normal modes, i.e., $\alpha_{\mathbf{q}l}(t)=\alpha_{\mathbf{q}l}e^{-i\omega_{\mathbf{q}l}t}$.
If  time evolution is ignored, then $\alpha_{\mathbf{q}l}(t)=\alpha_{\mathbf{q}l},$ which is equivalent to setting the frequencies to zero $\omega_{\mathbf{q}l}=0$ (except for those appearing in $g_{\mathbf{q}l}$ and $N_{\mathbf{q}l}$).
Thus, the inverse momentum relaxation time is 
\begin{align}
    \ev{\frac{1}{\tau_{tr}^{(C_0)}}}
    &=
    -\sum_{\mathbf{q}}
    \frac{\mathbf{k}\cdot\mathbf{q}}{\mathbf{k}^2}
    \frac{2\pi g_{\mathbf{q}l}^2}{\hbar}
    2N_{\mathbf{q}l}
    \delta(\varepsilon(\mathbf{k+q})-\varepsilon(\mathbf{k}))|_{\varepsilon(\mathbf{k})=E_F}
    \label{eq:tautrthCFTI}
\end{align}

\end{widetext}
which shows there is no phonon absorption or emission (cf. the classical time-dependent field in Eq. \eqref{invTauCF}).
Necessarily, the scattering in the frozen field is elastic \cite{davies_physics_1998} unlike in a time-dependent field.

The results from the different fields in Eqs. \eqref{invTauIEScattInteg2}, \eqref{eq:tautrthCF}, and \eqref{eq:tautrthCFTI} can be condensed to the following equation:
\begin{align}
    \ev{\frac{1}{\tau_{tr}^{(F)}}}
    &=
    -\sum_{\mathbf{q}}
    \frac{\mathbf{k}\cdot\mathbf{q}}{\mathbf{k}^2}
    \frac{2\pi g_{\mathbf{q}l}^2}{\hbar}
    2N_{\mathbf{q}l}J_{\mathbf{q}l}^{(F)}
    \notag
    \\
    &\qquad\qquad\times\delta[\varepsilon(\mathbf{k+q})-\varepsilon(\mathbf{k})]|_{\varepsilon(\mathbf{k})=E_F},
    \label{eq:tautrall}
\end{align}
where $F$ is an index indicating which field (quantum field $F=Q$, classical field $F=C$, and classical time-independent field $F=C_0$) is used and $J_{\mathbf{q}l}^{(F)}$ is a factor appearing in the corresponding field choice
\begin{align*}
    J_{\mathbf{q}l}^{(Q)}&=(N_{\mathbf{q}l}+1)\beta\hbar\omega_{\mathbf{q}l},
    \\
    J_{\mathbf{q}l}^{(C)}&=(N_{\mathbf{q}l}+1/2)\beta\hbar\omega_{\mathbf{q}l},
    \\
    J_{\mathbf{q}l}^{(C_0)}&=1.
\end{align*}
Note $J_{\mathbf{q}l}^{(Q)}>J_{\mathbf{q}l}^{(C)}>J_{\mathbf{q}l}^{(C_0)}=1$ for $\beta\hbar\omega_{\mathbf{q}l}>0$. In the high-temperature limit, where all the phonon energies are small compared to thermal energy, all the factors become   unity $\lim\limits_{\beta\hbar\omega_{\mathbf{q}l}\to0}J_{\mathbf{q}l}^{(Q)}=\lim\limits_{\beta\hbar\omega_{\mathbf{q}l}\to0}J_{\mathbf{q}l}^{(C)}=J_{\mathbf{q}l}^{(C_0)}=1$, so there is no difference between the choice of fields in the high-temperature limit, within   perturbation theory.

The underestimation of the inverse momentum relaxation time in the classical fields is significant  near or below the critical temperature $T_c=\min\{T_D,T_{\mathrm{BG}}\}$ as shown in Fig. \ref{InvTauFields}.
The inset of Fig. \ref{InvTauFields} shows the ratios of the inverse momentum relaxation times to the quantum field value.
The difference between the blue and red curves is from the missing spontaneous emission, which suppresses the rate by about a factor of 2.
The difference between the red and green curves arises from ignoring the time dependence of the potential (elastic approximation), which, with the missing spontaneous emission, decreases the rate by five (four) times in 3D (2D). The temperature dependence is, however, correctly captured with the classical fields,  allowing introduction of  a correction factor to reach the proper values, for example based on Fig. \ref{InvTauFields}.

The explicit forms of Eq. \eqref{eq:tautrall} in two and three dimensions are
\begin{align}
    \frac{1}{\tau_{tr}^{(F,2D)}}
    &=
    \frac{m^*}{2\pi\hbar^3k_F^3}
    \int_{0}^{q_{\mathrm{max}}}\frac{\dd q q^2}{\sqrt{1-(q/2k)^2}}
    \dfrac{E_{d}^2\hbar q}{\rho_m v_s}
    \frac{J_{\mathbf{q}l}^{(F)}}
    {e^{\hbar \omega_{\mathbf {q}l}/k_{B}T}-1}
    \label{invTau2DDP}
\end{align}
\begin{align}
    \frac{1}{\tau_{tr}^{(F,3D)}}
    &=
    \frac{m^*}{4\pi\hbar^3k_F^3}
    \int_{0}^{q_{\mathrm{max}}}\dd q q^3
    \dfrac{E_{d}^2\hbar q}{\rho_m v_s}
    \frac{J_{\mathbf{q}l}^{(F)}}
    {e^{\hbar \omega_{\mathbf {q}l}/k_{B}T}-1}
    \label{invTau3DDP}
\end{align}

Note the quantum field gives the conventional results.
In 2D, from Eq. \eqref{invTauIEScattInteg2} we obtain
\begin{align}
    \ev{\frac{1}{\tau_{tr}^{(Q,2D)}}}
    &\approx
    \frac{m^*}{2\pi\hbar^3k_F^3}
    \int_{0}^{q_{\mathrm{max}}}\frac{\dd qq^2}{\sqrt{1-(q/2k_F)^2}}
    \frac{E_d^2\hbar q}{\rho_m v_s}\notag
    \\
    &\qquad\times
    N_{ql}(N_{ql}+1)\beta\hbar\omega_{ql}
    \label{invTau2DDPinelastic}
\end{align}
where $q_{\mathrm{max}}=\min\{q_D,2k\}=2k$, 
which is basically the formula derived and used by Hwang and Das Sarma~\cite{T4}, and Efetov and Kim~\cite{kim,pseudospinfactor}. To explicitly connect this to their resistivity result, use the relation $\Delta\rho^{\mathrm{(2D)}}(T)=\frac{m^*}{ne^2}\ev{\frac{1}{\tau_{tr}^{\mathrm{(2D)}}}}$ where $k_F^2=\pi n$, $q_{\mathrm{max}}=2k_F$ and $m^*v_F=\hbar k_F$ for graphene [they also considered absence of backscattering due to chiral nature of the carriers
in graphene, which leads to additional factor of $1-(q/2k_F)^2$ in the integrand in Eq. \eqref{invTau2DDPinelastic}].

In 3D, from Eq. \eqref{invTauIEScattInteg2} we obtain
\begin{align}
    \ev{\frac{1}{\tau_{tr}^{(Q,3D)}}}
    &=
    \frac{m^*}{4\pi\hbar^3k_F^3}
    \int_{0}^{q_{\mathrm{max}}}\dd q q^3
    \frac{E_d^2\hbar q}{\rho_m v_s}
    N_{ql}(N_{ql}+1)\beta\hbar\omega_{ql}
    \label{invTau3DDPinelastic}
\end{align}
which is basically the Bloch--Gr\"{u}neisen formula~\cite{ziman2001electrons}. To explicitly connect this to the resistivity result in the citation, use the relation $\Delta\rho^{\mathrm{(3D)}}(T)=\frac{m^*}{ne^2}\ev{\frac{1}{\tau_{tr}^{\mathrm{(3D)}}}}$ where $n=\frac{k_F^3}{3\pi^2}$, $q_{\mathrm{max}}=\min\{q_D,2k\}=q_D$ and $m^*v_F=\hbar k_F$ for normal metals.

\section{Semiclassical dynamics on the deformation potential: Ray trajectories }\label{app:SemiclassicalRay}
\subsection{Semiclassical dynamics on the deformation potential}
In the classic solid state textbook by Ashcroft and Mermin~\cite{ashcroft1976solid}, two chapters are devoted to ``semiclassical" methods, by which is meant treating external fields acting on electrons in metals as classical fields, with the kinetic part of the Hamiltonian governed by the band structure. 
Consider a metal at $\SI{0}{K}$. For a given electronic band structure $E_0(\mathbf{k})$, the electron group velocity is $ \dd\mathbf{r} /\dd t = \partial E_0(\mathbf{k})/\partial (\hbar\mathbf{k})$; $\mathbf{r}$ and $\hbar\mathbf{k}$ are single-electron position and momentum, respectively. If there are externally applied electric and magnetic fields $\bm{\mathcal{E}}$ and $\bm{\mathcal{B}}$, the traditional phenomenological semiclassical model gives
$\dd{(\hbar\mathbf{k})} /\dd t = -e   [\bm{\mathcal{E}}(\mathbf{r},t)  + \dot {\mathbf{r} }\times  \bm{\mathcal{B}}(\mathbf{r},t)]$~\cite{ashcroft1976solid, kittel}.
Consequently, electron motion is treated classically. This is  an {\it ad hoc} idea, but it works extremely well.

A crucial observation is that, for region III in Fig. \ref{fig:phasediaBoth}, electrons are in a semiclassical regime with respect to the deformation potential, having short Fermi wavelengths compared to the length scale of the deformation potential. This
supports the idea that the deformation potential $V_D(\mathbf{r},t)$ can be
considered as a classical field acting on the electrons, just
as external fields are treated. Then, Hamiltonian is given as $H(\hbar\mathbf{k},\mathbf{r},t)=E_0(\mathbf{k})+V_D(\mathbf{r},t)$.
We expect the (phenomenological) Hamilton's equations of motion to be
\begin{equation}
\label{apex}
     \dv{(\hbar\mathbf{k})}{t} = -\frac{\partial H(\mathbf{k}, \mathbf{r},t)}{\partial \mathbf{r}}; \ \ \ 
    \dv{\mathbf{r}}{t} = \ \frac{\partial H(\mathbf{k}, \mathbf{r},t)}{\partial (\hbar\mathbf{k})}.
\end{equation}
This allows use of nonperturbative semiclassical trajectory-based methods, just as
external electromagnetic fields are routinely treated~\cite{ashcroft1976solid, kittel}.

We numerically investigate the electron scattering by the deformation potential using a fourth-order symplectic scheme for integration~\cite{YoshidaSymplectic}.
For each temperature, we run thousands of trajectories with the same
initial kinetic energy $E_F$, using random initial directions and
positions, and several realizations of the random deformation potential. Then, average momentum relaxation time can be calculated from the simulation results.

\begin{figure}[t]
\centering
\includegraphics[width=0.48\textwidth]{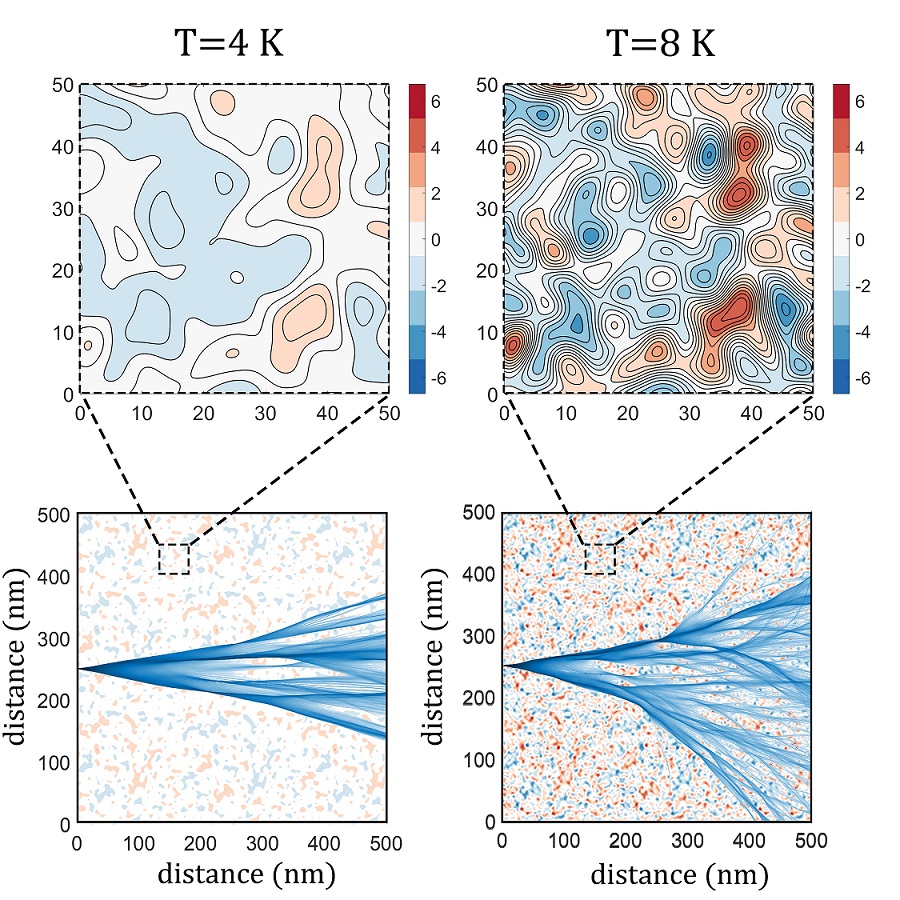}
\caption{Deformation potential and semiclassical ray trajectories at two different temperatures 4 and $\SI{8}{K}$ for graphene~\cite{kim}.
In blue tones, ensembles of classical electron pathways launched uniformly over a small range of angles revealing branched flow.
Identical random phases are used to generate the deformation potential for both temperatures.
It shows the emergence of new vibrational modes with increasing temperature.
The bumps and dips also get higher and deeper with increasing temperature.}
\label{semicl}
\end{figure}

\begin{figure}
\centering
\includegraphics[width=0.44\textwidth]{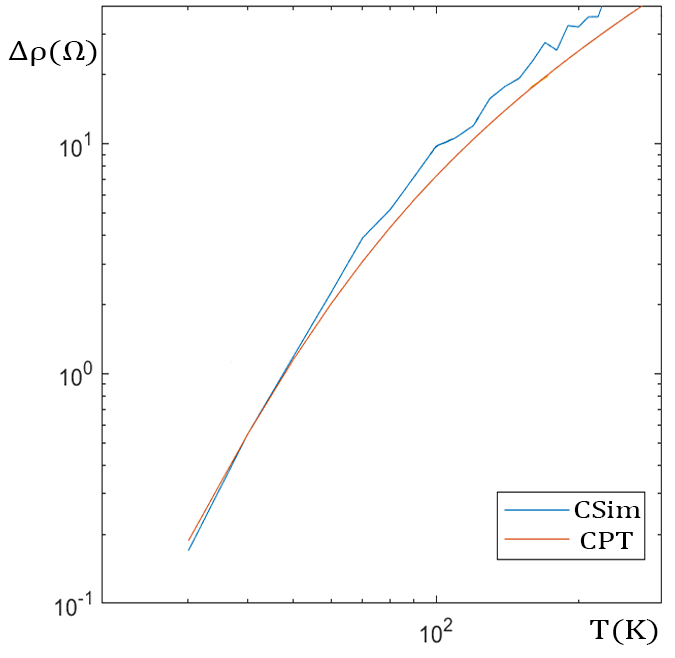}
\caption{Resistivity calculated from classical simulation (labeled as CSim, blue curve) and classical perturbation theory (labeled as CPT, red curve). The two curves match well since the parameter set is in a perturbative and classical regime.}
\label{CPT}
\end{figure}

The deformation potential and electron trajectories are shown in Fig. \ref{semicl}. On the top, the contour maps of the deformation potentials for two different temperatures are given. On the bottom, the electron trajectories launched from a point in space over a small range of angles are shown.
In this classical and perturbative regime (region III in Fig. \ref{fig:phasediaBoth}), electrons make small-angle forward scatterings under the deformation potential and they exhibit branched flow~\cite{topinka2001coherent,Heller2008, metzger2010universal,segev}. Electrons forming branched flow patterns have been experimentally shown in 2D electron gas at very low temperatures~\cite{topinka2001coherent}.
The forces from the semiclassical deformation potential on conduction electrons have the correct properties to explain electron deflection and pure metal Drude resistivity as checked by numerical results and perturbation theory as shown in Fig. \ref{CPT}. For region III in Fig. \ref{fig:phasediaBoth}, perturbative and classical regime, the classical simulations give consistent results with not only the classical perturbation theory, but also quantum simulation and perturbation theory given in the main text as explained in the next section.

\subsection{Classical perturbative derivation of the temperature dependence of resistivity}\label{app:PertClassical}
As the sound speed $v_s$ is far smaller than the Fermi velocity $v_F$ of an electron, it is reasonable to consider the frozen deformation potential $V_D(\mathbf{r},t)|_{t=0}$ for this purpose.
For the frozen deformation potential, the momentum correlation function can be calculated analytically using classical perturbation theory. One knows $\delta\mathbf{p}(t)=\mathbf{p}(t)-\mathbf{p}(0)=\int_0^t\dd t'\left(-\pdv{V_D(\mathbf{r}(t'))}{\mathbf{r}(t')}\right)$. Then, $\mathbf{p}(0)\cdot\mathbf{p}(t)=\abs{\mathbf{p}(0)}\abs{\mathbf{p}(t)}\cos\chi(t)\approx\abs{\mathbf{p}(0)}^2\left(1-\frac{\abs{\delta\mathbf{p}(t)}^2}{2\abs{\mathbf{p}(0)}^2}\right)$ where $\chi(t)$ is the angle between the two momenta, and the quasielasticity of scattering $\abs{\mathbf{p}(t)}\approx\abs{\mathbf{p}(0)}$.
Take the ensemble average (average over all possible realizations of deformation potentials specified by $\{\varphi_{\mathbf{q}}\}$) to obtain $\ev{c(t)}
    =
    \ev{\mathbf{p}(0)\cdot\mathbf{p}(t)}
    =
    \abs{\mathbf{p}(0)}^2\left(1-\frac{\ev{\abs{\delta\mathbf{p}(t)}^2}}{2\abs{\mathbf{p}(0)}^2}\right)$
where
$\ev{\abs{\delta\mathbf{p}(t)}^2}
    =
     E_d^2
    \sum\limits_{\substack{\Vec q\\\abs{\Vec q}<q_{\mathrm{max}}}}
    \frac{2\hbar q}{\rho_m \mathcal{V} v_s}
    \left(\frac{\mathbf{q}}{\mathbf {q}\cdot\mathbf{p}(0)/m^*}\right)^2
    \frac{1-\cos(\mathbf {q}\cdot\mathbf{p}(0)t/m^*)}
    {e^{\hbar v_s q/k_BT}-1}$
using unperturbed trajectory $\mathbf{r}^{(0)}(t')=\mathbf{p}(0)t'/m^*$. Note that the sinusoidal oscillation with different phases vanishes due to the ensemble average, \textit{i.e.}, $\ev{\cos(A+\varphi_{\mathbf{q}})\cos(B+\varphi_{\mathbf{q}'})}=\delta_{\mathbf{q},\mathbf{q}'}\ev{\cos(A+\varphi_{\mathbf{q}})\cos(B+\varphi_{\mathbf{q}})}$.

In the 2D case, we can use polar coordinate for $\mathbf{q}$ where angle $\theta$ is chosen such that $\mathbf {q}\cdot\mathbf{p}(0)=q\abs{\mathbf{p}(0)}\sin\theta$, and use $\abs{\mathbf{p}(0)}=m^*v_F$, then
\small
\begin{eqnarray*}
  \begin{aligned}
    \ev{\abs{\delta\mathbf{p}(t)}^2}
    &=
     E_d^2
    \int_0^{q_D}\frac{\mathcal{V}\dd q\;q}{(2\pi)^2}
    \int_0^{2\pi}\dd\theta
    \frac{2\hbar q}{\rho_m \mathcal{V} v_s}
    \left(\frac{1}{v_F\sin\theta}\right)^2
    \\
    &\qquad\qquad\qquad\qquad\qquad\times \frac{1-\cos(qv_Ft\sin\theta)}
    {e^{\hbar v_s q/k_BT}-1}
    \\
    &=
     E_d^2
    \int_0^{q_D}\frac{\dd q\;q}{(2\pi)^2}
    \frac{2\hbar q}{\rho_m v_s}
    \frac{1}{v_F^2}
    \frac{f(qv_Ft)}
    {e^{\hbar v_s q/k_BT}-1}
\end{aligned}
  \end{eqnarray*}
\normalsize
where $f(A)=A \pi \{J_1(A)[-2+A\pi H_0(A)]+A J_0(A)[2-\pi H_1(A)]\}$ and $H$'s and $J$'s are Struve and Bessel functions, respectively. Looking at sufficiently long time correlation such that $A=qv_Ft\gg1$ holds, we can approximate $f(A)\approx 2\pi A$.
Then, we obtain momentum relaxation time $\tau$ from $\ev{c(t)}=\abs{\mathbf{p}(0)}^2(1-t/\tau)$ for $t\ll\tau$, and obtain inverse momentum relaxation time
\begin{align}
    \frac{1}{\tau_{cl}^{\mathrm{(2D)}}}
    =
    \frac{m^*}{2\pi\hbar^3k_F^3} \int_0^{q_D}\dd qq^2\frac{E_d^2\hbar q}{\rho_m v_s}\frac{ 1}{e^{\hbar v_s q/k_BT}-1}
    \label{invTauCL2D}
\end{align}
where the subscript $cl$ stands for ``classical,'' and used $\abs{\mathbf{p}(0)}=m^*v_F=\hbar k_F$. Equation \eqref{invTauCL2D} does not have a geometrical factor $\sqrt{1-(q/2k_F)^2}$ appearing in the quantum elastic perturbation theory in Eq. \eqref{invTau2DDP}, but other than that the other parts of the integrand are the same. This shows classical-quantum correspondence. 

Likewise, in 3D, we obtain
\begin{align}
    \frac{1}{\tau_{cl}^{\mathrm{(3D)}}}
    &=
    \frac{m^*}{4\pi\hbar^3k^3}
    \int_{0}^{q_D}\dd q q^3
    \dfrac{E_{d}^2\hbar q}{\rho_m v_s}
    \frac{1}
    {e^{\hbar \omega_{\mathbf {q}l}/k_{B}T}-1}
    \label{invTauCL3D}
\end{align}
Equation \eqref{invTauCL3D} has the same integrand as the quantum elastic perturbation theory in Eq. \eqref{invTau3DDP}, showing classical-quantum correspondence.
However, note that the integration range of classical results in Eqs. \eqref{invTauCL2D} and \eqref{invTauCL3D} is up to $q_D$, in contrast to $q_{\mathrm{max}}$ in quantum results in Eqs. \eqref{invTau2DDP} and \eqref{invTau3DDP}. This means that, unlike in quantum mechanics, there is no ``transparency'' to $q_{\mathrm{max}}<q<q_D$ components of the potential in classical mechanics.
This difference will not be problematic if one uses classical simulation only in semiclassical regime $2k_F\ll q_D$ where $q_{\mathrm{max}}=q_D$.

\section{Chosen parameter values}\label{app:param}
\subsection{A typical metal}\label{app:paramnormal}
We chose parameter values from a typical metal: copper~\cite{ashcroft1976solid}. As there is no 2D copper, one needs to generate mock 2D parameters from 3D parameters. From the 3D mass density $\rho_m^{\mathrm{(3D)}}=\SI{8940}{kg.m^{-3}}$ and the radius of a copper atom $r_{\mathrm{Cu}}=\SI{1.35e-10}{m}$, the thinnest layer of copper ``sheet'' would roughly have a mass density 
$\rho_m^{\mathrm{(2D)}}=\rho_m^{\mathrm{(3D)}}2r_{\mathrm{Cu}}=\SI{2.41e-6}{kg.m^{-2}}$ (\textit{cf.} the mass density of graphene $\SI{7.6e-7}{kg.m^{-2}}$ is in the similar order).
For copper, sound speed is $v_s=\SI{4.7e3}{m.s^{-1}}$, the deformation potential constant is roughly $E_d=\SI{10}{eV}$~\cite{DPconstCu,DPconstCu2}, Debye temperature is $T_D=\SI{343}{K}$, Fermi energy is $E_F=\SI{7}{eV}$. 

\subsection{A strange metal}\label{app:paramstrange}

We chose YBCO as a material for strange metal parameters. The sound velocity is $v_s=\SI{6.3e3}{m.s^{-1}}$~\cite{PhysRevB.100.241114}, the deformation potential constant is chosen to be $E_d=\SI{20}{eV}$\cite{hwangSarma2019}, Debye temperature is $T_D=\SI{400}{K}$~\cite{ikebe}, and the 2D mass density is $\rho_m=\SI{3.6e-6}{kg.m^{-2}}$. The carrier (hole) density is $n=(1+p)/a^2$~\cite{Badoux2016} where $a=\SI{0.38}{nm}$ is in-plane lattice constant. Effective mass is $m^*=3m_e$~\cite{PhysRevB.72.060511} where $m_e$ is bare electron mass.
The charge carrier (hole) energy at the Fermi surface is calculated using $E_F^{(h)}=\hbar^2\pi n/m^*=\SI{0.64}{eV}$~\cite{PhysRevB.100.241114,PhysRevB.62.3554} at the critical doping level $p=0.16$~\cite{Badoux2016,Stangl2021}.

\bibliography{refs}

\begin{thebibliography}{140}
\bibitem [{ran()}]{randvars}%
  \BibitemOpen
  \href@noop {} {}\bibinfo {howpublished} {For any integer $n$, the position,
  time, and disorder averages of $(V_D)^n$ are \begin{align*}
  \ev{(V_D)^n}_{\mathbf{r}}&=\mathcal{V}^{-1}\int_{\mathcal{V}}\dd\mathbf{r}(V_D(\mathbf{r},t;\{\varphi_{\mathbf{q}l}\}))^{n},
  \\ \ev{(V_D)^n}_{t}&=\mathcal{T}^{-1}\int_0^\mathcal{T}\dd
  t(V_D(\mathbf{r},t;\{\varphi_{\mathbf{q}l}\}))^{n}, \\
  \ev{(V_D)^n}_{dis.}&=\left(\sum_{\{\varphi_{\mathbf{q}l}\}}1\right)^{-1}\sum_{\{\varphi_{\mathbf{q}l}\}}(V_D(\mathbf{r},t;\{\varphi_{\mathbf{q}l}\}))^{n},
  \end{align*} respectively, and they give the identical results, i.e.,
  $\ev{(V_D)^n}_{\mathbf{r}}=\ev{(V_D)^n}_{t}=\ev{(V_D)^n}_{dis.}$ as the
  position $\mathbf{r}$, time $t$, and phase $\varphi_{\mathbf{q}l}$ appear in
  the argument of the same cosine in Eq. \eqref{eq:VDcl}.}\BibitemShut {Stop}%
\end{thebibliography}

\end{document}